\documentclass[twocolumn]{aastex63}
\usepackage{amsmath}

\usepackage{xcolor, fontawesome, microtype}

\usepackage{ mathrsfs }
\usepackage[T1]{fontenc}
\usepackage{xcolor}

\received{\today}
\revised{\today}
\accepted{\today}
\submitjournal{AJ}

\shorttitle{K2: The Rising Phoenix}

\shortauthors{Zink et al. (2021)}

\begin{document}

\title{Scaling \emph{K2}. IV. A Uniform Planet Sample for Campaigns 1-8 and 10-18}

\correspondingauthor{Jon Zink}
\email{jzink@astro.ucla.edu}

\author[0000-0003-1848-2063]{Jon K. Zink}
\affiliation{Department of Physics and Astronomy, University of California, Los Angeles, CA 90095}

\author[0000-0003-3702-0382]{Kevin K. Hardegree-Ullman}
\affiliation{Caltech/IPAC-NASA Exoplanet Science Institute, Pasadena, CA 91125}

\author[0000-0002-8035-4778]{Jessie L. Christiansen}
\affiliation{Caltech/IPAC-NASA Exoplanet Science Institute, Pasadena, CA 91125}

\author[0000-0002-6673-8206]{Sakhee Bhure}
\altaffiliation{Volunteer Researcher}
\affiliation{Caltech/IPAC-NASA Exoplanet Science Institute, Pasadena, CA 91125}

\author{Britt Duffy Adkins}
\affiliation{Sol Price School of Public Policy, University of Southern California, Los Angeles, CA 90089}

\author[0000-0003-0967-2893]{Erik A. Petigura}
\affiliation{Department of Physics and Astronomy, University of California, Los Angeles, CA 90095}

\author[0000-0001-8189-0233]{Courtney D. Dressing}
\affiliation{Department of Astronomy, University of California, Berkeley, CA 94720}

\author[0000-0002-1835-1891]{Ian J. M. Crossfield}
\affiliation{Department of Physics and Astronomy, University of Kansas, Lawrence, KS 66045}

\author[0000-0001-5347-7062]{Joshua E. Schlieder}
\affiliation{Exoplanets and Stellar Astrophysics Laboratory, Code 667, NASA Goddard Space Flight Center, Greenbelt, MD 20771}



\begin{abstract}

We provide the first full \emph{K2} transiting exoplanet sample, using photometry from Campaigns 1-8 and 10-18, derived through an entirely automated procedure. This homogeneous planet candidate catalog is a crucial to perform a robust demographic analysis of transiting exoplanets with \emph{K2}. We identify 747 unique planet candidates and 57 multi-planet systems. Of these candidates, 366 have not been previously identified, including one resonant multi-planet system and one system with two short-period gas giants. By automating the construction of this list, measurements of sample biases (completeness and reliability) can be quantified. We carried out a light curve-level injection/recovery test of artificial transit signals and found a maximum completeness of 61\%, a consequence of the significant detrending required for \emph{K2} data analysis. Through this operation we attained measurements of the detection efficiency as a function of signal strength, enabling future population analysis using this sample. We assessed the reliability of our planet sample by testing our vetting software {\tt EDI-Vetter} against inverted transit-free light curves. We estimate 91\% of our planet candidates are real astrophysical signals, increasing up to 94\% when limited to the FGKM dwarf stellar population. We also constrain the contamination rate from background eclipsing binaries to less than $5\%$. The presented catalog, along with the completeness and reliability measurements, enable robust exoplanet demographic studies to be carried out across the fields observed by the \emph{K2} mission for the first time.

\end{abstract}



\section{Introduction}
The \emph{Kepler} space telescope collected continuous photometric data for nearly 3.5 years from a small 0.25\% patch of the celestial sphere \citep{koc10,bor16}. This uniform monitoring of 150,000 stars enabled the discovery of over 4700 transiting exoplanet candidates through an automated detection pipeline \citep{jen10}.\footnote{\url{https://exoplanetarchive.ipac.caltech.edu/docs/counts_detail.html}} Removing the human component from the signal vetting process ({\tt Robovetter}; \citealt{tho18}), enabled the first homogeneous transiting small planet sample suitable for exoplanet population studies. Through full automation, the sample completeness can be measured via transit injection/recovery procedures \citep{pet13b,chr15,dres15,chr17,chr20}. Here, artificial transit signals are injected into the raw light curves, enabling quantification of the pipeline's detection efficiency as a function of, for instance, the injection signal strength. In addition, the final \emph{Kepler} catalog (DR25) also provided a measure of the sample reliability (the rate of false alarm signals) by testing the software's ability to remove systematics from the final catalog \citep{cou17b}.   

With these catalog measurements available, the underlying planet population can be extracted by correcting for the detection and orbital selection effects. Many studies have used \emph{Kepler} automated planet catalogs to identify a remarkable surplus of sub-Neptunes and super-Earths at short periods (FGK Dwarfs: \citealt{you11,how12,pet13b,muld18,hsu19,zin19,he19}; M Dwarfs: \citealt{dres13,mui15,dres15,har19}), despite their absence in the solar system. Several other population features have been discovered using the \emph{Kepler} catalog. For example, the apparent deficit of planets near the 1.5--$2 R_\Earth$ range, colloquially known as the ``radius valley'' \citep{ful17}. This flux-dependent population feature indicates some underlying formation or evolution mechanism must be at play, separating the super-Earth and sub-Neptune populations \citep{owe17,gup18}. Nearly 500 multi-planet systems were identified in the \emph{Kepler} data, enabling examination of intra-system trends. Remarkably, very minor dispersion in planet radii is observed within each system \citep{cia13}. By examining the intra-system mass values from planetary systems that have measurable transit-timing variations (TTVs), a similar uniformity has been identified for planet mass \citep{mil17}. Moreover, the orbital period spacing of planets appears far more compact than the planets within the solar system. These transiting multi-planet systems also appear to have relatively homogeneous spacing, indicative of an unvaried dynamic history. This combination of intra-system uniformity in orbital spacing and radii is known as the "peas-in-a-pod" finding \citep{wei18}.

One of the main objectives of the \emph{Kepler} mission was to provide a baseline measurement for the occurrence of Earth analogs ($\eta_\Earth$). Despite the scarce completeness in this area of parameter space, several studies have extrapolated from more populated regions, providing measurements ranging from 0.1--0.4 Earth-like planets per main-sequence star \citep{cat11,tra12,pet13b,sil15,bur15,zin19b,kun20,bry20b}. Currently, it remains unclear what planet features are important for habitability, making precise measurements unattainable. Moreover, these extrapolations are model dependent and require a more empirical sampling of the Earth-like region of parameter space to reduce our uncertainty. Unfortunately, the continuous photometric data collection of the \emph{Kepler} field was terminated due to mechanical issues on-board the spacecraft after 3.5 years, limiting the completeness of these long-period small planets.

Upon the failure of two (out of four) reaction wheels on the spacecraft, the telescope was no longer able to collect data from the \emph{Kepler} field, due to drift from solar radiation pressure. By focusing on fields (Campaigns) along the ecliptic plane, this drift was minimized, giving rise to the \emph{K2} mission \citep{how14, cle16}. Each of the 18 Campaigns were observed for roughly 80 days, enabling the analysis of transiting exoplanet populations from different regions of the local Galaxy. However, the remaining solar pressure experienced by the spacecraft still required a telescope pointing adjustment every $\sim$6 hours. This drift and correction led to unique systematic features in the light curves, which the \emph{Kepler} automated detection pipeline was not designed to overcome. As a result, all transiting \emph{K2} planet candidates to date have required some amount of visual assessment. Despite this barrier, nearly 1000 candidates have been identified in the \emph{K2} light curves (e.g., \citealt{van16b,bar16,ada16,cro16,pop16,dres17,pet18,liv18,may18,ye18,kru19,zin19c}). However, this assortment of planet detections lacks the homogeneity necessary for demographics research. Attempts have been made to replicate the automation of the \emph{Kepler} pipeline for \emph{K2} planet detection \citep[e.g.,][]{kos19,dat19}, but these procedures still required some amount of visual inspection. \citet{zin20a} developed the first fully automated \emph{K2} planet detection pipeline, using the {\tt EDI-Vetter} suite of vetting metrics to combat the unique \emph{K2} systematics. This pilot study was carried out on a single \emph{K2} field (Campaign 5; C5 henceforth) and detected 75 planet candidates. The automation enabled injection/recovery analysis of the stellar sample, providing an assessment of the planet catalog completeness. Additionally, the reliability (rate of false alarms) was quantified by passing inverted light curves---nullifying existing transit signals---through the automated procedure, analogous to what was done for the \emph{Kepler} DR25 catalog. Any false candidates identified through this procedure are indicative of the underlying false alarm contamination rate. 

\begin{figure*}
\centering \includegraphics[width=\textwidth]{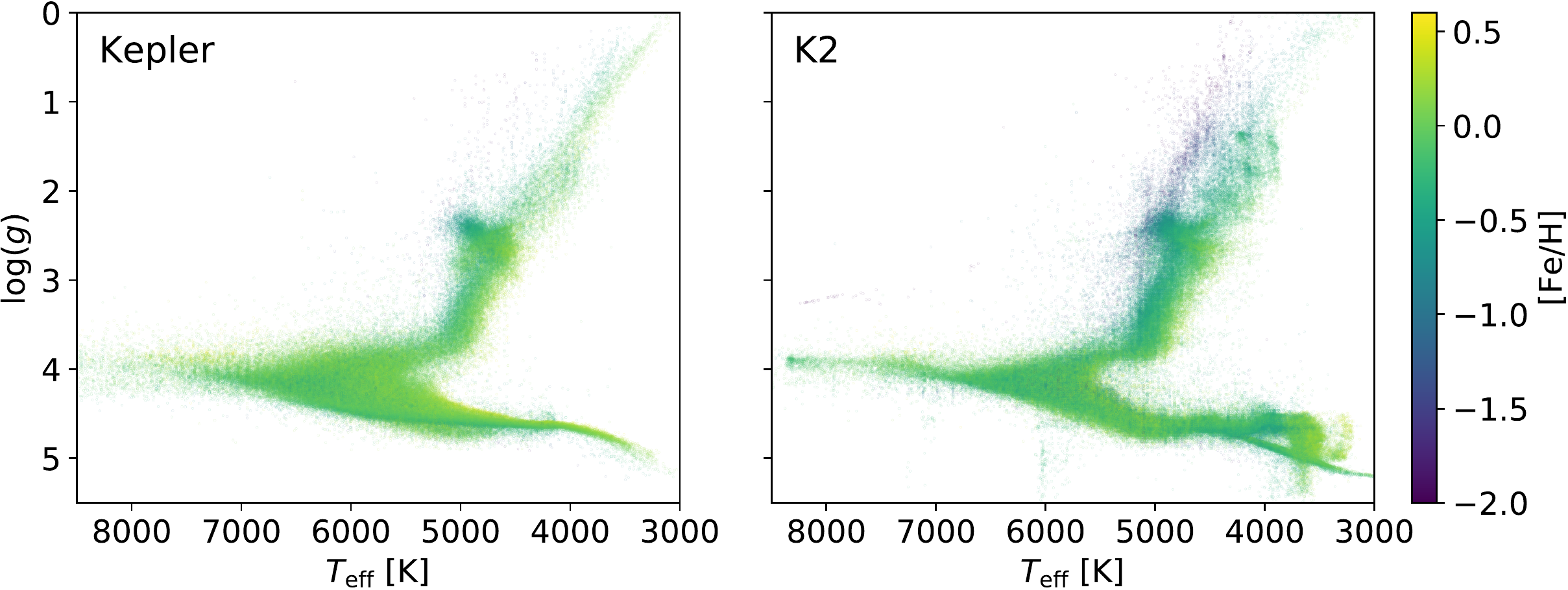}
\caption{The distribution of \emph{Kepler} \citep{ber20} and \emph{K2} \citep{hub16,har20} targets as a function of surface gravity ($\log(g)$), effective stellar temperature ($T_\textrm{eff}$) and stellar metallicity ([Fe/H]). 
\label{fig:HR}}
\end{figure*}

With corresponding measures of sample completeness and reliability for C5, \cite{zin20b} carried out the first assessment of small transiting planet occurrence outside of the \emph{Kepler} field, finding a minor reduction in planet occurrence in this metal-poor FGK stellar sample. This provided evidence that stellar metallicity may be linked to the formation of small planets; however, the weak trend detected requires additional data for verification.

This paper is a continuation of the Scaling \emph{K2} series \citep{har20,zin20a,zin20b}, which aims to leverage \emph{K2} photometry to expand our exoplanet occurrence rate capabilities and disentangle underlying formation mechanisms. The intent of this study is to derive a uniform catalog of \emph{K2} planets for all 18 Campaigns (with exception to C9, which observed the crowded galactic plane, and C19, which suffered significantly from low fuel levels), and provide corresponding measurements of the sample completeness and reliability. The underlying target sample and the corresponding stellar properties are discussed in Section \ref{sec:stellarSample}. 
In Section \ref{sec:pipe} we outline the automated pipeline implemented in our uniform planet detection routine. We then procure a homogeneous planet sample and consider the interesting candidates and systems in Section \ref{sec:catalog}. In Sections \ref{sec:complete} and \ref{sec:reli} we provide the corresponding measures of sample completeness and reliability for this catalog of transiting \emph{K2} planet candidates. Finally, we offer suggestions for occurrence analysis and summarize our findings in Sections \ref{sec:suggest} and \ref{sec:summary}.

\section{Target Sample}
\label{sec:stellarSample}

We first downloaded the most up-to-date raw target pixel files (TPFs) from MAST\footnote{\href{https://archive.stsci.edu/k2/}{https://archive.stsci.edu/k2/}}, which recently underwent a full reprocessing using a uniform procedure \citep{cad20}. To ensure consistency among our data set, we used the final data release (V9.3) for all available campaigns considered in this paper.\footnote{C8, C12, and C14 were released after our analysis was performed. Thus, a previous version (V9.2) was used for this catalog. Overall, we expect the differences to be minor and leave inclusion of these updated TPFs for future iterations of the catalog.} We began our search using the entire EPIC target list (381,923 targets; \citealt{hub16}). Of this list, we found 212 targets have FITS file issues our software could not rectify. For all remaining targets we used the {\tt K2SFF} aperture \#15 \citep{van16b}, which is derived from the \emph{Kepler} pixel response function and varies in size in accordance with the target's brightness. Circular apertures exceeding $10\arcsec$ radii are prone to significant contamination from nearby sources. Our vetting software, {\tt EDI-Vetter}, is able to account for this additional flux, but the software's accuracy begins to decay beyond a radius of $20\arcsec$. Additionally, large apertures have a tendency to contain multiple bright sources, providing further complications. To address this issue, we put an upper limit on the target aperture size (79 pixels; $\sim20\arcsec$ radius). This cut removed 4548 targets that have apertures we deemed too large. Of these rejected targets, 492 are in Campaign 11, which contains the crowded galactic bulge field. After applying these cuts, 377,163 targets remained and were then used as our baseline stellar sample.

\begin{figure*}
\centering \includegraphics[height=7cm]{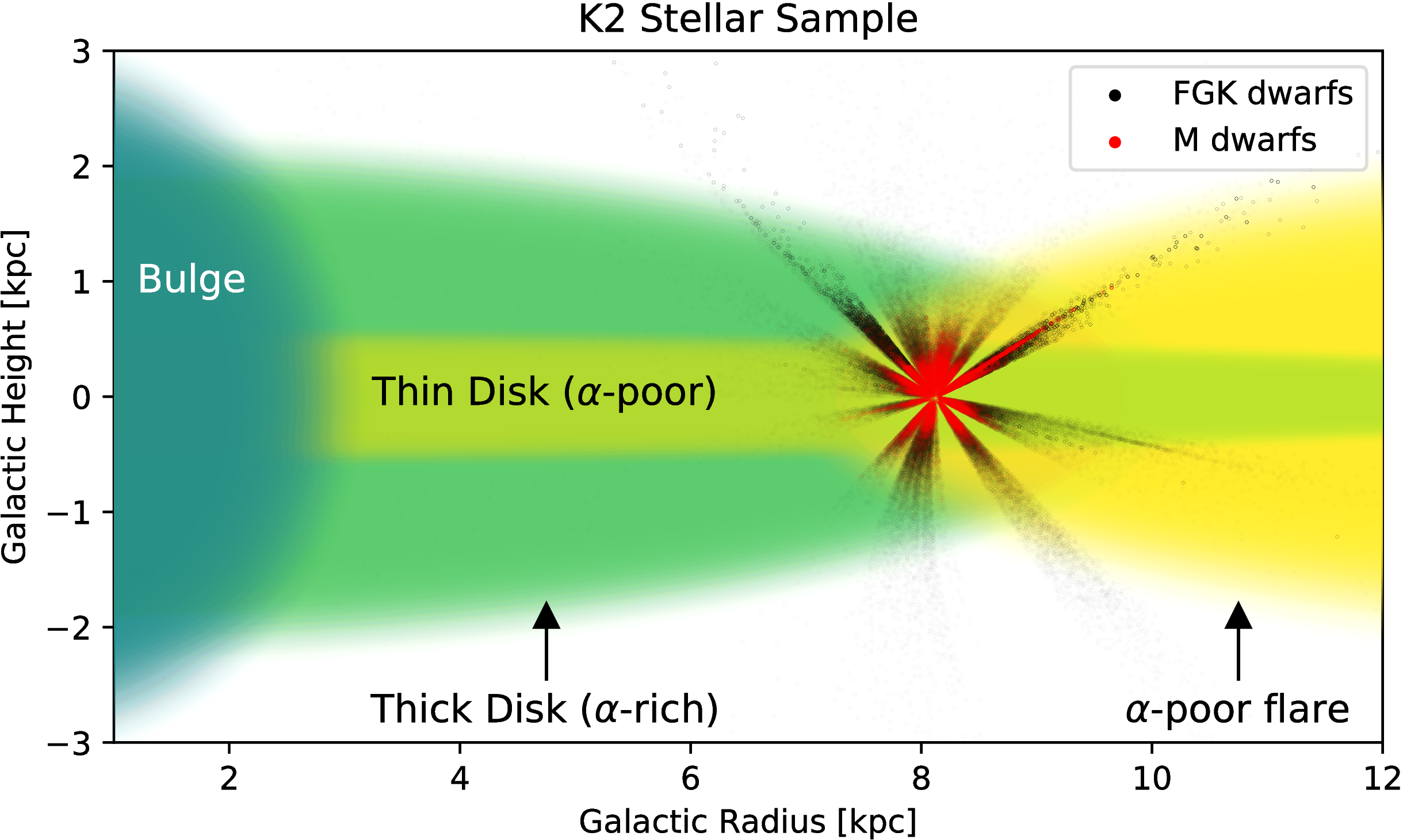}
\caption{The galactic distribution of \emph{K2} target stars. The disk structure follows the interpretation provided by \cite{hay15}. The thick disk consists of stars with a heightened abundance of $\alpha$-chain elements (O, Ne, Mg, Si, S, Ar, Ca, and Ti) compared to the stars in the thin disk \citep{wal62}. Additionally, the $\alpha$-poor disk begins to flare up beyond the solar neighborhood. The galactic coordinates presented here have been calculated using \emph{Gaia} DR2 \citep{gai18}.
\label{fig:galaxy}}
\end{figure*}

To parameterize this list of targets we relied on the values derived by \cite{har20}, which used a random forest classifier, trained on LAMOST spectra \citep{su04}, to derive stellar parameters from photometric data (222,088 unique targets). Additionally, the available \emph{Gaia} DR2 parallax information was incorporated to significantly improve our understanding of the stellar radii. From this catalog, additional measurements of stellar mass, metallicity, effective temperature ($T_{\text{eff}}$), and surface gravity ($\log(g)$) were provided and used in this study. In cases where stellar parameters were not available in \cite{har20} (typically due to their absence in the \emph{Gaia} DR2 catalog or not meeting strict photometric selection criteria), we used the stellar parameters derived for the EPIC catalog \citep{hub16} (94,769 targets). Finally, we used solar values for targets that lack parameterization in both catalogs (41,061 targets).\footnote{\cite{zin20b} showed that systematic differences between stellar catalogs are minor. Additionally, we emphasize our pipeline's agnosticism to stellar parameterization.}

Excluding the 41,061 targets without stellar parameters, we isolated 223,075 stars that appear to be main-sequence dwarfs based on their surface gravity ($\log(g)>4$). Within this subset we identified 48,702 targets as M dwarfs ($T_{\text{eff}}<4000K$), 164,569 as FGK dwarfs ($4000<T_{\text{eff}}<6500K$), and 9,731 as A stars ($T_{\text{eff}}>6500K$). This abundance of M dwarfs is nearly 17 times that of the \emph{Kepler} sample (2808 M dwarfs; \citealt{ber20}) and can be clearly seen in Figure \ref{fig:HR}. Another noteworthy feature of the \emph{K2} stellar sample is the wide range of galactic latitudes covered by the 18 campaigns. This enabled \emph{K2} to probe different regions of the galactic sub-structure. In contrast, a majority of the \emph{Kepler} stellar sample was bounded within the thin disk (see Figure 1 of \citealt{zin20b} for a comparison). In Figure \ref{fig:galaxy} we show this galactic sub-structure span for \emph{K2}. Making broad cuts in galactic radius ($R_g$) and height ($b$) we can distinguish thick disk ($R_g<8$kpc and $|b|>0.5$kpc) from thin disk stars ($|b|<0.5$kpc). Overall, we found 191,002 dwarfs located in the thin disk, while 17,123 dwarfs reside in the thick disk. This sample distinction is interesting and may warrant further consideration in future occurrence studies.

The underlying stellar sample is important because it is the population from which the planet candidates are drawn. However, the stellar parameters are subject to change as more comprehensive data and more precise measurements become available (i.e., the upcoming \emph{Gaia} DR3). To ensure our catalog remains relevant upon improved stellar parameterization, we take an agnostic approach to the available stellar features throughout our pipeline. In other words, each light curve is treated consistently regardless of the underlying target parameters. The only caveat to this claim is our treatment of the transit limb-darkening. Our transit model fitting routine requires quadratic limb-darkening parameters. We derived the appropriate values using the ATLAS model coefficients for the \emph{Kepler} bandpasses \citep{cla12}, in concert with the available stellar parameters. This minor reliance on our measured stellar values will have negligible effects on the presented catalog. In the case where a signal is transiting with a high impact parameter, this limb darkening choice can be important, modifying the inferred radius ratio by an order of 1\%. However, this boundary case is well within the uncertainty of our measured radius ratio values ($\sim4\%$) and therefore not significantly impacting our inferred radius measurements. All other detection and vetting metrics in our pipeline are independent of this parameterization.

\section{The Automated Pipeline}
\label{sec:pipe}
\begin{figure}
\centering \includegraphics[width=\columnwidth{}]{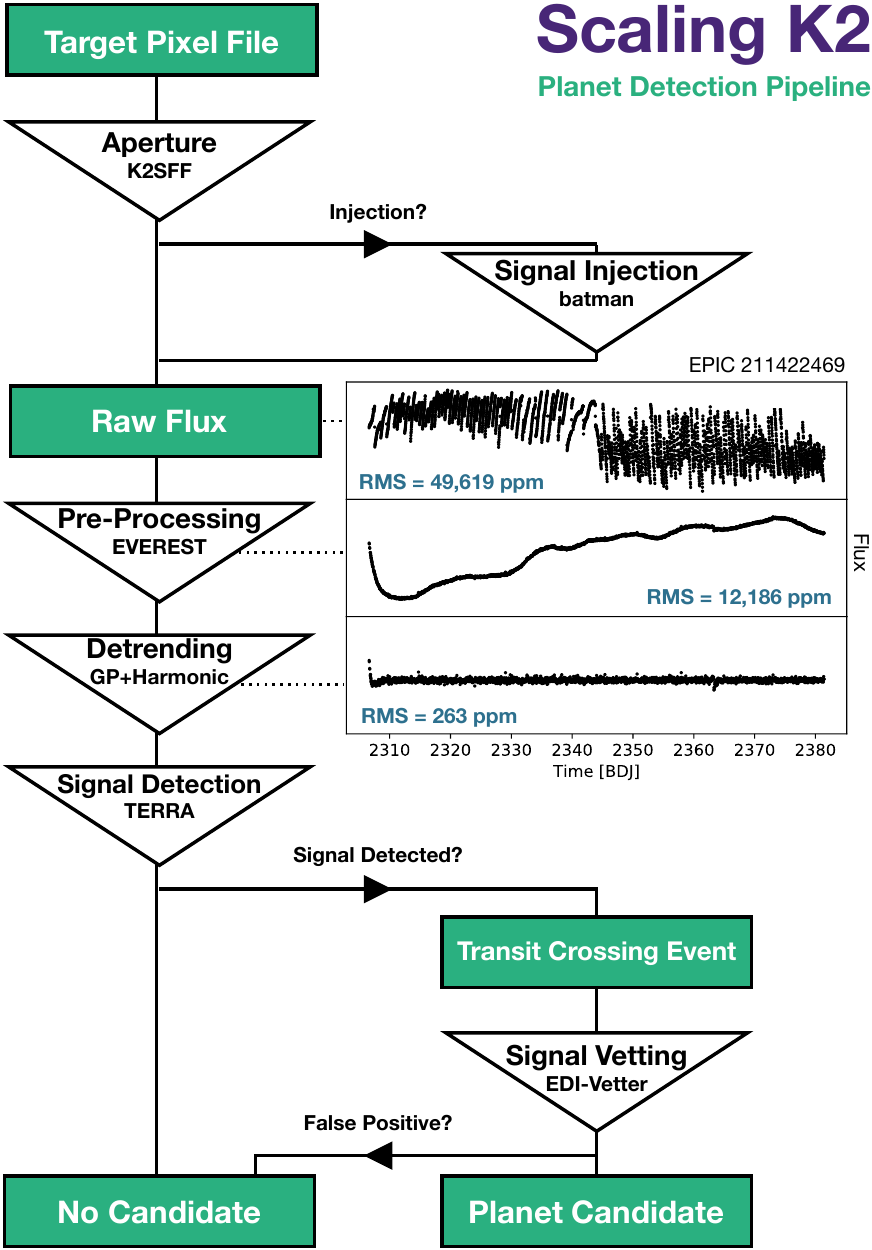}
\caption{ An overview schematic of the automated detection pipeline used in the current study to identify \emph{K2} planet candidates. This follows the same procedure described in \cite{zin20a}. The effects of pre-processing and detrending have also been depicted for EPIC 211422469, illustrating the importance of the corresponding steps. 
\label{fig:diagram}}
\end{figure}

Our automated light curve analysis pipeline consists of four major components: pre-processing, detrending, signal detection, and signal vetting. In Figure \ref{fig:diagram} we provide a visual overview of this procedure, and now briefly summarize the execution of each step.

\subsection{Pre-Processing}
The raw flux measurements from \emph{K2} are riddled with systematic noise components due to the thruster firing approximately every six hours (required to re-align the telescope pointing) and the momentum dumps every two days. This spacecraft movement smears the target across several different pixels, all with unique noise and sensitivity properties, leading to a significant increase in the overall light curve noise. We passed all raw light curves through the {\tt EVEREST} software \citep{lug16,lug18}, which minimizes this flux dispersion issue using pixel-level decorrelation (PLD) to fit and remove noise attributed to the spacecraft roll. In Figure \ref{fig:diagram} we show how {\tt EVEREST} reduces the noise, as measured by the root mean-square (RMS), of the EPIC 211422469 light curve by a factor of four. However, this pre-processing also has the ability to reduce transit signals or remove them completely. Thus, it is essential that any injection/recovery tests address this concern by injecting signals into the data before this pre-processing, as we later discuss in Section \ref{sec:complete}.

\subsection{Detrending}
In the example light curve for EPIC 211422469 (Figure \ref{fig:diagram}) there is a clear long-term trend remaining in the data after pre-processing. The goal of detrending is to remove this red-noise component and any stellar variability, producing a clean, white-noise dominated, time series. We used two Gaussian process (GP) models, with ``rotation'' kernels, to remove these long- and medium-term trends. These kernels use a series of harmonics to match and remove periodic and red-noise trends in the photometry.  The first pass GP looked for general flux drifting (periods $>10$ days), subtracting the appropriate model. The second pass GP identified and removed medium-term trends (5 days $<$ period $<10$ days) often associated with stellar variability. In testing, we found these two GP passes were effective in removing red-noise from the data without significantly impacting transit signals. The Ljung-Box test (a portmanteau test for all autocorrelation lags; \citealt{lju78}) found that $67\%$ of our processed light curve residuals produced p-values greater than 0.001, indicating a lack of statistically significant short-term correlated structure. However, stellar harmonics with short periods (usually $<0.5$ days) continued to contaminate the remaining 33\% of our light curves. To address these trends we fit and removed sine waves with periods less than $0.5$ days. These signals are all below the 0.5 period requirement threshold for planet candidacy, but still have the ability to contaminate the light curve. In an effort to minimize over-fitting, we required the signal amplitude to be at least $10\sigma$ in strength. If this requirement was not met, the harmonic removal was not applied.  

The period limits used for both the GP models and the harmonic fitter attempted to preserve transit signals, avoiding periods which are prone to transit removal (0.5 days $<$ period $<5$ days). Unfortunately, some stellar variability exists in this forbidden period range, making it difficult to address every unique situation. Using an auto-regressive integrated moving average algorithm (as suggested by \citet{cac19a,cac19b}) may reduce some of these remaining correlated residuals, but this methodology also has limitations in dealing with stellar variability and is beyond the scope of this paper.

In addition, these detrending mechanisms are capable of reducing or (in the case of very deep transits) removing the signal altogether (i.e., the \emph{Kepler} harmonic removal highlighted by \citealt{chr13}).We acknowledge these costs in performing our automated detrending procedure, knowing the effects will be accounted for in our catalog completeness measurements (see Section \ref{sec:complete}).  

\subsection{Signal Detection}
Once the light curve has been scrubbed with our pre-processing and detrending routine, the flux measurements can be examined for transit signals with {\tt TERRA} \citep{pet13b}. This algorithm uses a box shape to look for dips in the light curve, enabling a quick examination of each target. To measure the signal strength we rely on the same metric used for the \emph{Kepler} transiting planet search (TPS), the multiple event statistic \citep[MES;][]{je02}, which assumes a linear ephemeris to indicate the strength of the whitened signal. For our detection threshold we require a signal MES value greater than $8.68\sigma$. This threshold, which is higher than that of the \emph{Kepler} TPS ($7.1\sigma$), was arbitrarily selected to reduce the total signal count to 20\% of our total target sample. We also bound the period search of {\tt TERRA}, ranging from 0.5 days to a period where three transits could be detected (nominally 40 days, varying slightly from campaign to campaign). Any signal above our MES threshold and between these period ranges is given the label of threshold-crossing event (TCE).

\subsubsection{Multi-planet systems}

Multi-planet systems are rich in information, but require careful consideration when attempting to identify these asynchronous signals. Ideally, a model of the first detected signal would be fit and subtracted from the light curve before re-searching the data for additional signals. However, real data are noisy, making such a task difficult to automate. Should the signal be incorrectly fit, the model subtraction will leave significant residuals, which will continuously trigger a detection upon reexamination. Moreover, false positives that do not fit the transit model will also leave significant residuals. Finally, astrophysical transit timing variations (TTVs) are not easily accounted for in an automated routine \citep{wu13,had14,had17}. Thus, deviations from simple periodicity will leave significant residuals. Dealing with these complex signals, without loss of data, remains a point of continuous discussion (e.g. \citealt{sch17}).

As was done in many previous pipelines, we relied on masking of the signal after each iterative TCE detection \citep{jen02,dres15,sin16,kru19,zin20a}. This method has its own faults, as it requires some photometry to be discarded upon each signal detection. \citet{zin19} showed that this $3\times$ the transit duration removal has the ability to make multi-planet systems more difficult to detect. To mitigate this loss of data, we used a mask of $2.5\times$ the transit duration ($1.25\times$ the transit duration on either side of the transit midpoint) after each signal was detected. At this point the light curve was reexamined for an additional signal. This process was repeated six times or until the light curve lacked a TCE, enabling this pipeline to detect up to six planet systems.

\subsubsection{Skye Excess TCE Identification}

\begin{figure}
\centering \includegraphics[width=\columnwidth{}]{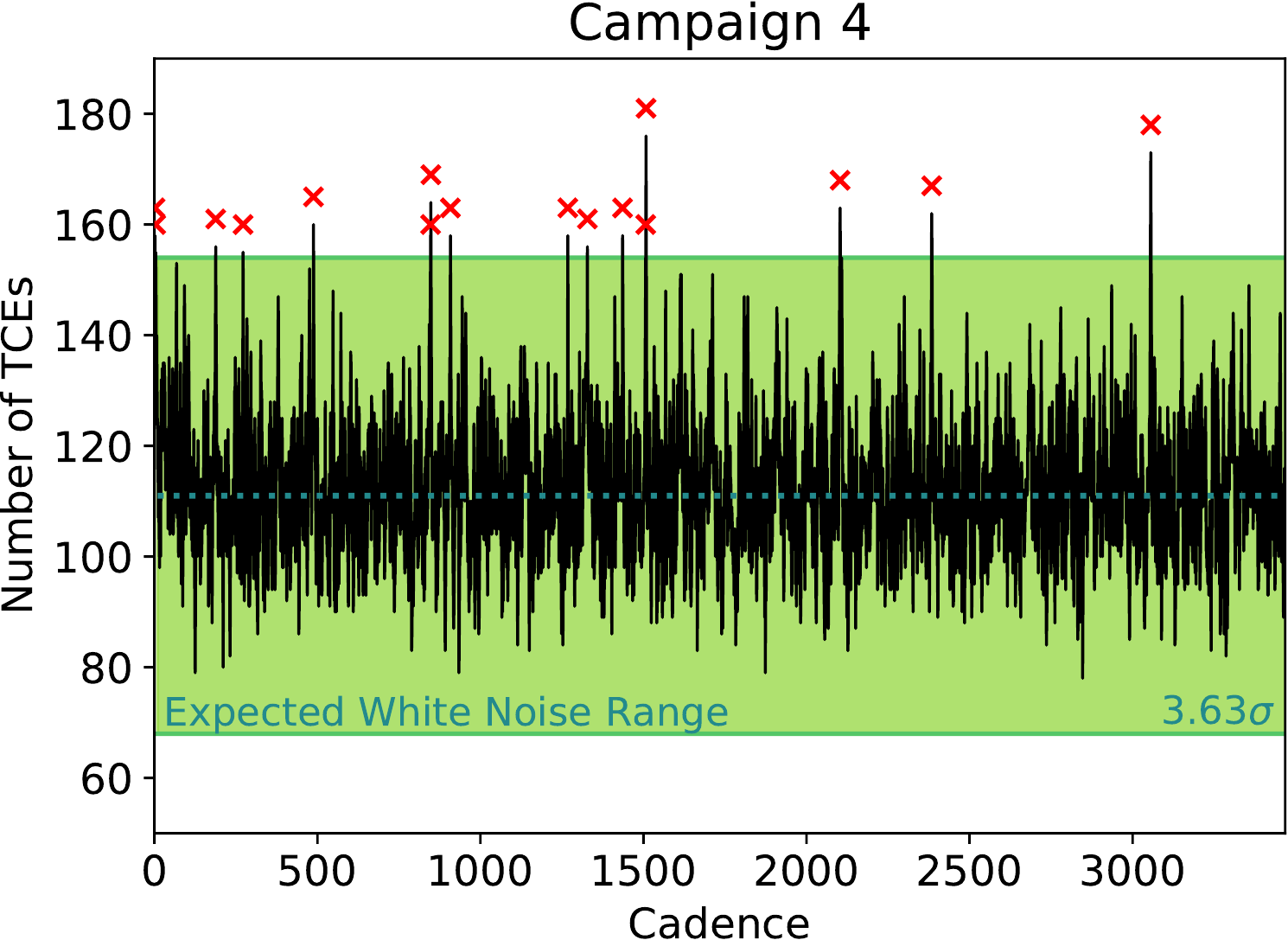}
\caption{ An illustration of the Skye excess TCE identification process for Campaign 4. The expected white noise range for this number of cadences is plotted in green ($3.63\sigma$). The median number of TCEs is shown by a central blue dotted line. We highlight the rejected cadences with a red x. These 16 cadences exceed the Skye excess limit and are masked in our final signal search.  
\label{fig:sky}}
\end{figure}

Certain cadences are prone to triggering TCEs within the population of light curves. These can usually be attributed to spacecraft issues and are likely not astrophysical. Inspired by the ``Skye'' metric used in \citet{tho18}, we minimized this source of contamination by considering the total number of TCEs with transits that fall on each cadence. The median value and expected white noise range was then calculated using:

\begin{equation}
\sqrt{2}\; \text{erfcinv}(1/N_{\text{cad}}) \; \sigma,
\label{eq:sky}
\end{equation}
where erfcinv is the inverse complementary error function, $N_{\text{cad}}$ is the total number of cadences in the campaign, and $\sigma$ is the median absolute deviation in the number of TCEs detected at each cadence. This calculated value represents the largest deviation expected from this number of cadences under the assumption of perfect Gaussian noise. Any cadences that exceed this limit are likely faulty and warrant masking. In Figure \ref{fig:sky} we provide an example of this procedure for Campaign 4. To further assist future studies we made the Skye mask for each campaign publicly available.\footnote{\href{http://www.jonzink.com/scalingk2.html}{http://www.jonzink.com/scalingk2.html}}

Once established, the Skye mask was used to reanalyze all light curves, removing problematic cadences. The resulting list of significant signals were then given the official TCE label and allowed to continue through our pipeline. Overall, we found 140,046 TCEs from 52,192 targets. However, a vast majority of these signals are false positives and required thorough vetting.

\subsection{Signal Vetting}
To parse through the 140,046 TCEs and identify the real planet candidates, we employed our vetting software {\tt EDI-Vetter} \citep{zin20a}. This routine builds upon the metrics developed for the \emph{Kepler} TPS ({\tt RoboVetter}; \citealt{tho18}), with additional diagnostics created to address \emph{K2} specific issues (i.e., the systematics caused by the spacecraft). These metrics attempt to replicate human vetting, by looking for specific transit features used to discern false positive signals. We now briefly discuss our planet candidacy requirements, but encourage readers to refer to \citet{zin20a} for a more thorough discussion.

\subsubsection{Previous Planet Check}
This test looks to identify duplicate signals in the light curve (as originally discussed in Section 3.2.2 of \citealt{cou16}). This test was only applied if the light curve produced more than one TCE. In such cases, the period and ephemeris were tested to ensure the second signal was truly unique. If not, this repeat identification was labeled a false positive. The goal of this test was to remove detections of the previous signal's secondary eclipse and signals that were not properly masked, leading to their re-detection.

\subsubsection{Binary Blending}
\label{sec:bblend}
The goal of this metric is to remove eclipsing binary (EB) contaminants from our catalog. There are two attributes that make EB events unique from planet transits: a deep transit depth and a high impact parameter ($b$). Leveraging these two features, we used the formula first derived by \citet{bat13} and then modified by \cite{tho18} to identify these contaminants, while remaining agnostic to the underlying stellar parameters:

\begin{equation}
\frac{R_{pl}}{R_\star} + b\le 1.04,
\label{eq:EB}
\end{equation}
where $R_{pl}/R_{\star}$ represents the ratio of the transiting planet and the stellar host radii. Should this metric be exceeded, the TCE would be flagged as a false positive, removing most EBs from our sample.

However, this diagnostic assumes the transit depth provides an accurate measure of $R_{pl}/R_{\star}$. If an additional source is within the target photometric aperture, the depth of the transit can be diluted by this additional flux, leading to an underestimation of $R_{pl}/R_{\star}$. To address this potential flux blending issue, we cross-referenced our target list against the \emph{Gaia} DR2 catalog \citep{gai18}. If an additional source is in or near the aperture, we calculated the expected flux contamination using available photometry from \emph{Gaia} and \emph{2MASS} \citep{skr06}, correcting the $R_{pl}/R_{\star}$ value accordingly. If Equation \ref{eq:EB} was not satisfied, the TCE was labeled as a false positive.

Our inferred flux dilution correction assumed the signal in question originated from the brightest source encased by the target aperture. \footnote{This assumption was enforced by our pipeline, which selected the brightest \emph{Gaia} target within the aperture.} Therefore, transits emanating from a dimmer background star would experience a greater flux dilution and could avert this metric. We provide further discussion of this potential contamination rate in Section \ref{sec:astrFP}.

\subsubsection{Transit Outliers}
This diagnostic was developed to deal with systematics specific to \emph{K2}. We expect real candidates to produce dips in the stellar flux, but retain comparable light curve noise properties during the eclipse. In contrast, systematic events can produce a dip with noise properties independent of the light curve, producing a heightened RMS during the event. By measuring the RMS in and out of transit, we can identify significant changes, and flag signals that have false positive-like RMS properties.

\subsubsection{Individual Transit Check}
\label{sec:ITC}
Using the formalism developed by \citet{mul16} (the ``Marshall'' test), we look to identify individual transit events that appear problematic. If one of the transits is either dominating the signal strength or does not fit a transit profile appropriately, it is indicative of a systematic false alarm. By fitting each individual transit signals with any of the four common systematic models (Flat, Logistic, Logistic-exponential, or Double-logistic), we looked for events that did not have an astrophysical origin. If an individual transit fit a systematic model better, that transit was masked and the light curve was re-analyzed to ensure the signal MES still remained above the $8.68\sigma$ threshold before proceeding. In addition, we made sure the total signal strength is spread evenly among each observed transit. In cases where a single event dominated the signal strength, the TCE was labeled as a false positive.  

\subsubsection{Even/Odd Transit Test}
This vetting test looks for EBs that produce a strong secondary eclipse (SE). In some cases these SEs can be deep enough to trigger a signal detection at half the true period, folding the SE on top of the transit. To identify such contaminants, we separated every other individual transit into odd and even groups. We then re-fit the transit depth within each group and looked for significant discrepancies. In cases where the disparity is greater than $5\sigma$, the signal was labeled as a false positive.

\subsubsection{Uniqueness Test}
This diagnostic is based on the ``model-shift uniqueness test'' \citep{row15, mul15, cou16, tho18} and compares the noise profile of the folded light curve with that of the TCE in question. If the folded light curve contains several transit-like features, it is indicative of a systematic false alarm. We compared the strength of the next largest dip, beyond the initial signal and any potential secondary eclipse, in the phase-folded data to the transit in question, assessing the uniqueness and significance of the periodic event. If another similar magnitude dip existed in the folded time series, it likely originated from light curve noise and the TCE was deemed a false positive.

\subsubsection{Check for Secondary Eclipse}
Following the methodology presented in Section A.4.1.2 and A.4.1.3 of \citet{tho18}, we examined the transit signal for a secondary eclipse (SE). As previously noted, deep SE signals are a notable signature of an EB. However, some hot Jupiters are also capable of producing a SE, therefore detection of a SE is not in itself justification for exclusion. TCEs with a meaningful SE must exhibit a transit impact parameter of less than 0.8 or the SE must be less than 10\% of the transit depth. If neither of these criteria were satisfied, the TCE was labeled as a false positive.

\subsubsection{Ephemeris Wandering}
Harmonic signals have the ability to falsely trigger a TCE detection, but their inability to match the transit model leads to significant movement of the measured ephemeris. If the signal's transit mid-point, as detected by {\tt TERRA}, changed by more than half the transit duration when optimized with our MCMC routine, the TCE earned a false positive label. 

\subsubsection{Harmonic Test}
In addition to the ephemeris wandering metric, we also attempted to fit the light curve with a sine wave at the period of the detected TCE. If the amplitude of this harmonic signal was comparable to the TCE depth, or the strength of the harmonic signal was greater than $50\sigma$, the TCE received a false positive label. To avoid misclassification due to period aliasing, we examined periods of $2\times$ and $\frac{1}{2}\times$ the TCE. Additionally, harmonic signals tend to trigger TCEs with long transit durations, which often correspond to the sine wave period. Therefore, we also tested harmonics with periods equaling $1\times$, $2\times$, and $3\times$ the transit duration. If any of the examined phases exceeded our harmonic metric threshold, the TCE achieved a false positive label.

\subsubsection{Phase Coverage Test}
It is important that the transit signal has good phase coverage with the available \emph{K2} data. In certain cases, a few outlier points can be folded, near an integer multiple of the \emph{K2} cadence, and trigger a TCE. These limited data signals are not meaningful and warrant a false positive label. In addition, the masking applied by our automated pipeline may remove large fractions of the TCE. To ensure we have good signal phase coverage, we examined the gap sizes in the phase folded light curve. If large portions (more than roughly 30 mins.) of photometry are missing during ingress or egress, we labeled these signals as false positives.

\subsubsection{Period and Transit Duration Limits}
We limit the {\tt TERRA} software to period ranges beyond 0.5 days. However, this search algorithm is still able to identify periodic signals just below this user defined threshold.\footnote{This phenomenon is due to the method in which {\tt TERRA} steps through period space, moving in cadence integers until it exceeds the period limits.} These boundary case signals are usually astrophysical false positives and are difficult to distinguish from real planet candidates. Thus, we remove any TCEs with periods less than 0.5 days. Furthermore, we imposed a strict limit on the transit duration. If the TCE transit duration was greater than 10\% of the period of the signal, we deemed it a false positive, as these signals are often misidentified harmonics. Moreover, a transiting signal with a duration that exceeds this limit would represent a planet orbiting within three times  the radius of the host star for a short period eclipsing binary. Thus, removing these signals improves the purity of our sample.

\subsubsection{Period Alias Check}
Mis-folding a real transiting signal on an integer multiple of the true signal period will trigger many of the previously described false positive flags. To avoid this misclassification we tested period factors of $2\times$, $\frac{1}{2}\times$, $3\times$, and $\frac{1}{3}\times$ against the original signal. If any of the alternative models produced a likelihood value greater than 1.05 times the original period fit, we reran the entire vetting analysis, using the new corrected period.

\subsubsection{Ephemeris Match}
Deep transit signals from bright sources (the parent) can pollute neighboring target apertures (the child) and produce transit artifacts. These signals can be identified by their nearly identical period and ephemeris measurements. We followed the procedures of \citet{cou14b} and Section A.6 of \citet{tho18} to identify these false positives, testing our candidate list against itself and our full TCE sample, ensuring previously rejected deep eclipsing binary signals were considered.

With an ephemeris match it is important to deduce the signal's true origin, since both the parent and child targets will be identified by the matching algorithm. It is expected that the child signal will be significantly polluted by stray starlight, reducing the expected transit depth. Thus, we assign parenthood to the target with the largest transit depth signal and label all additional matches as false positives.

Overall, the implementation of the ephemeris match vetting metric only reduced our catalog by $\sim2\%$, considerably less than the $\sim6\%$ reduction found for the \emph{Kepler} DR25. This discrepancy can be attributed to the \emph{Kepler} field's higher target density, which was dominated by faint stars that are more susceptible to this type of false positive. Therefore, the increased average brightness of \emph{K2} targets and the reduced target density, explain this reduction in ephemeris matches. 

\begin{figure*}
\centering \includegraphics[height=6.5cm]{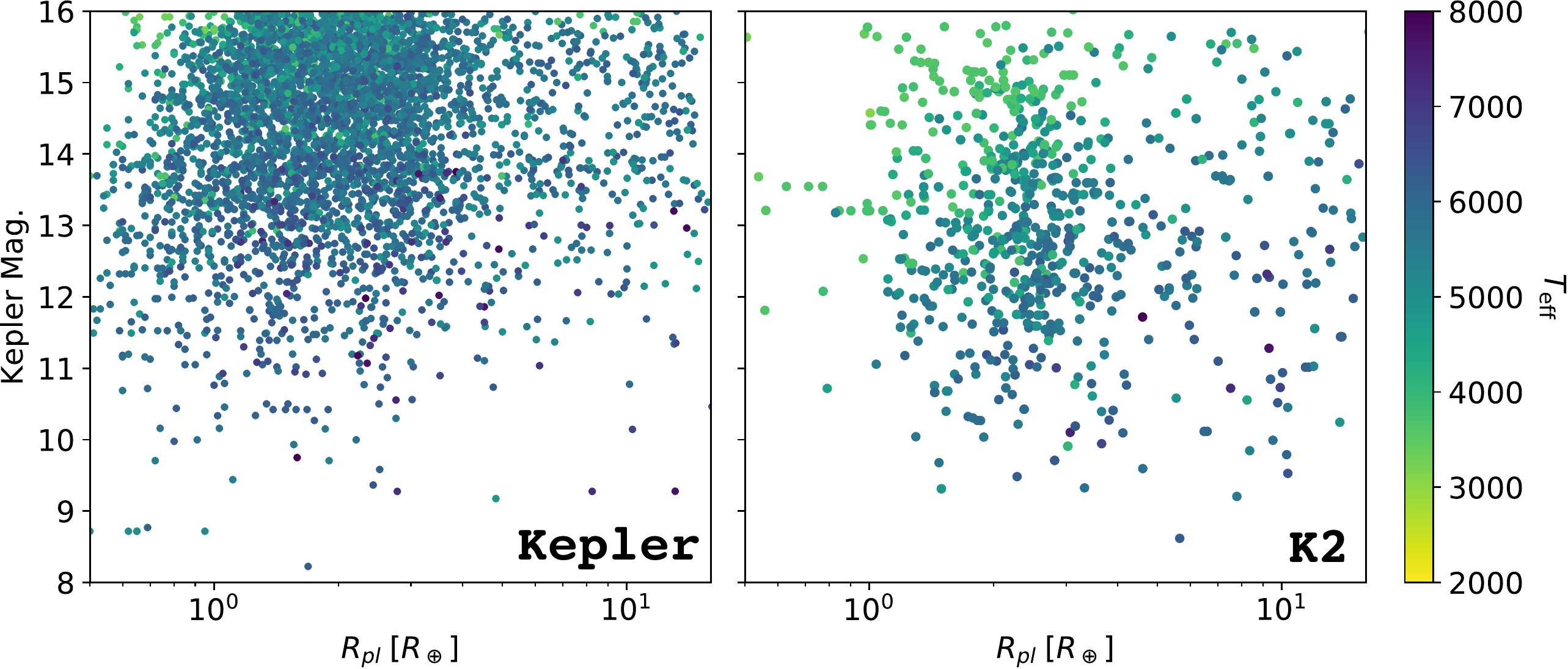}
\caption{The brightness and planet radius distributions of the \emph{Kepler} and \emph{K2} host stars, colored by host star effective temperature. The left panel shows the \emph{Kepler} candidates as described in \cite{ber20} and the right panel shows the catalog presented in this paper. \emph{K2} targets tend to be slightly brighter---with larger planets, due to the reduced completeness---than \emph{Kepler} targets, making them better candidates for follow-up surveys. 
\label{fig:host}}
\end{figure*}

\subsubsection{Consistency Score}
Finally, if the TCE passed all of the described vetting diagnostics without achieving a false positive flag, the light curve was reanalyzed---including detrending and signal detection---50 times. This final test measures the stochastic nature of the detrending and the MCMC parameter estimation. Any signal near one of the vetting thresholds would likely be pushed over during these reexaminations. The consistency score was then calculated as the number of times a given TCE was able to pass all the vetting metrics over the number of times tested. Any TCE able to achieve a consistency score greater than 50\% was granted planet candidacy. Of the 1046 TCEs that initially passed all the {\tt EDI-Vetter} thresholds, 806 met this final consistency requirement.

\section{The Planet Catalog}
\label{sec:catalog}
Using our fully automated pipeline, we achieved a sample of \emph{K2} transiting planets suitable for demographics. Within this catalog we found:
\begin{itemize}
    \item 806 transit signals
    \item 747 unique planet candidates
    \item 57 multi-planet candidate systems (113 candidates)
    \item 366 newly detected planet candidates
    \item 18 newly identified multi-planet candidate systems (38 candidates).
\end{itemize}
The majority of new candidates were found in campaigns not exhaustively searched (C10-18), with a few new candidates identified with low MES in early campaigns. Of the 806 transiting signals identified, 51 signals were detected in multiple, overlapping, campaigns. In the subsequent sections we discuss how the parameters of the planets were derived and the unique systems and candidates found in this catalog.

\subsection{Planet Parameterization}
\begin{figure}
\centering \includegraphics[width=\columnwidth{}]{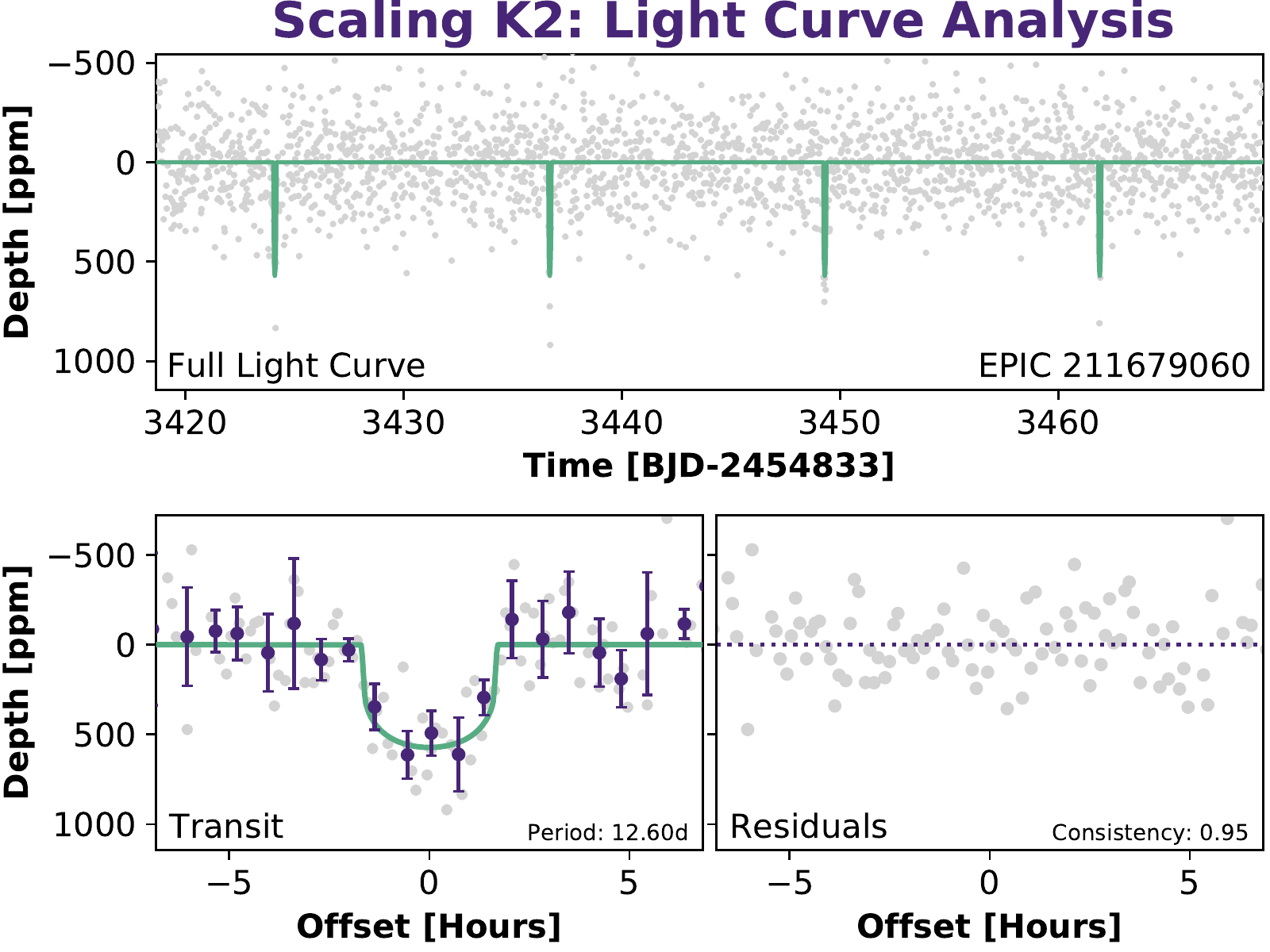}
\caption{A sample diagnostic plot for EPIC 211679060.01; the remaining candidate plots are available \href{http://www.jonzink.com/scalingk2.html}{online}. This planet is a new sub-Neptune found in C18. The grey points represent the light curve data used to extract the signal. The purple points show the binned average (with a bin width equal to 1/6 the transit duration).    \label{fig:newExample}}
\end{figure}

To maintain the homogeneity of our sample, we used an automated routine to fit and estimate the planet and orbital features. As discussed in Section \ref{sec:stellarSample} we assume the stellar parameters derived by \citet{har20}, but found 130 of the 747 detected planets did not have stellar characterization available. Fortunately, our pipeline is agnostic to changes in stellar features, enabling subsequent parameter revision. To avoid the assumption of solar values and provide the most current stellar parameters, we used the \citet{har20} methodology, equipped with APASS (DR9; \citealt{hen16}) and SkyMapper \citep{onk19} photometry, to characterize 59 of the remaining host stars. The final 71 targets, which did not have the necessary photometric coverage and/or parallax measurements for the aforementioned classification, were parameterized using the {\tt iscochrones} stellar modeling package \citep{mor15},\footnote{In order to maintain stellar parameter uniformity we ran the other 628 planet hosts through {\tt isochrones}. We fit a linear offset on these targets between our parameters and the {\tt isochrones} parameters, and applied these offsets to the 71 {\tt isochrones}-only targets.} ensuring all planets in our catalog have corresponding stellar measurements. Using the {\tt emcee} software package \citep{goo10, for13}, we measured the posterior distribution of the transiting planet parameters: The ephemeris, the radius ratio, the transit impact parameter, the period, the semi-major axis to stellar radius ratio (apl), the transit duration (tdur), and the vetting consistency score (score). These values were all derived under the assumption of circular orbits. Furthermore, we provide diagnostic plots for each candidate. In Figure \ref{fig:newExample} we show an example plot for a new sub-Neptune (EPIC 211679060.01), enabling swift visual inspection. However, it is important to clarify that all planets in our sample have been detected through our fully automated pipeline. These plots are only meant to help prioritize follow-up efforts. Figure \ref{fig:host} shows how the distribution of \emph{K2} host targets is skewed toward brighter stars, compared to the \emph{Kepler} candidate hosts, enabling follow-up efforts for a majority of our catalog.

With careful consideration of the transit radius ratio ($R_{fit}$), the planet radius can be extracted. $R_{fit}$ is directly measured by the MCMC routine, but estimates of the true planet radii ($R_{pl}$) required we take into account the stellar radii measurements ($R_{\star}$) and potential contamination from nearby sources ($\frac{F_{total}}{F_{\star}}$). To attain our best estimate we used the following procedure:
\begin{equation}
R_{pl} =R_{\textrm{fit}}\; R_{\star}\; \sqrt{\frac{F_{\textrm{total}}}{F_{\star}}},\\
\end{equation}
 where $\frac{F_{total}}{F_{\star}}$ was calculated using the \emph{Gaia} DR2 stellar catalog to identify contaminants in and near the photometric aperture and a Gaussian point-source function was implemented to estimate the corresponding magnitude of contamination. \footnote{We do not impose an upper $R_{pl}$ limit in our catalog, retaining our stellar parameter agnosticism. Therefore, 28 candidates exceed $30R_{pl}$. We provide suggestions for dealing with these candidates in Section \ref{sec:summary}.} In addition to the aforementioned flux complications, \cite{zin20a} also showed that the required detrending of \emph{K2} photometry underestimates the radius ratio by a median value of 2.3\%. We did not adjust our estimates of the planet radius to reflect this tendency, as the mode of this distribution indicated a majority of the measured radii ratios are accurate (see Figure 14 of \citealt{zin20a}). However, we increased the uncertainty in our planet radius measurements to account for this additional detrending complication. The overall planet radius uncertainty ($\sigma_R$) was calculated by assuming parameter independence and adding all of these relevant factors in quadrature,
\begin{equation}
\begin{aligned}[t]
\sigma_R &=  \sqrt{\sigma_{\textrm{fit}}^2+\sigma_{\star}^2+\sigma_F^2+\sigma_{\textrm{Off}}^2} \;,&  \textrm{where}\\
\sigma_{\textrm{Off}} & =  0.023\; R_{pl} & \textrm{and}\\
\sigma_F & =  R_{pl}\;\frac{F_{\textrm{total}}-F_{\star}}{3.76\; F_{\textrm{total}}}\;. &
\end{aligned}
\end{equation}
Here, $\sigma_{\textrm{Off}}$ and $\sigma_F$ represent the uncertainty due to detrending and flux contamination respectively (see Section 8.2 of \citealt{zin20a} for a thorough explanation and derivation of these parameters). Despite these additional contributions, the majority of uncertainty stems from the radius ratio fit ($\sigma_{\textrm{fit}}$; $\sim4\%$) and the stellar radius measurement ($\sigma_{\star}$; $\sim6\%$). For most targets $\sigma_F$ contributes of order $10^{-3}\%$ uncertainty, however, the most extreme candidate (EPIC 247384685.01) had a measured $F_{\star}$ that is only 70\% of $F_{\textrm{total}}$, contributing an additional $7\%$ uncertainty to the radius measurements.

\begin{deluxetable*}{lcc}
\tablecaption{The homogeneous catalog of \emph{K2} planet candidates and their associated planet and stellar parameters. The visual inspection flags were manually assigned to help prioritize follow-up efforts. These indicators had no impact on the analysis performed by this pipeline. The known planet (KP) flag indicates a confirmed or validated planet. The planet candidate (PC) flag identifies an unconfirmed candidate and the low priority planet candidate (LPPC) flag designates a weak or more difficult to validate candidate. Finally, the false positive (FP) flag specifies candidates that are likely not planets.      \label{tab:catalog}}
\tablehead{\colhead{Column} & \colhead{Units} & \colhead{Explanation}} 
\startdata
1 & --- &      EPIC Identifier  \\
2 & --- &      Campaign   \\
3 & --- &      Candidate ID\\
4 & --- &      Found in Multiple Campaigns Flag   \\
5 & --- &     Consistency Score \\
6 & d   &     Orbital Period  \\
7 & d   &     Lower Uncertainty in Period  \\
8 & d   &     Upper Uncertainty in Period  \\
9 & --- &     Planetary to Stellar Radii Ratio  \\
10 & --- &     Lower Uncertainty in Ratio  \\
11 & --- &     Upper Uncertainty in Ratio  \\
12 & $R_\earth$ &     Planet radius ($R_{pl}$)  \\
13 & $R_\earth$ &    Lower Uncertainty in $R_{pl}$  \\
14 & $R_\earth$ &    Upper Uncertainty in $R_{pl}$  \\
15 & d    &    Transit ephemeris ($t_0$)  \\
16 & d    &    Lower Uncertainty in $t_0$  \\
17 & d    &    Upper Uncertainty in $t_0$  \\
18 & ---  &    Impact parameter (b)  \\
19 & ---  &    Lower Uncertainty in b  \\
20 & ---  &    Upper Uncertainty in b  \\
21 & ---  &    Semi-major Axis to Stellar Radii Ratio ($a/R_\star$)   \\
22 & ---  &    Lower Uncertainty in $a/R_\star$  \\
23 & ---  &    Upper Uncertainty in $a/R_\star$  \\
24 & d    &    Transit Duration  \\
25 & $R_\sun$  &   Stellar Radius ($R_\star$)   \\
26 & $R_\sun$  &   Lower Uncertainty in $R_\star$  \\
27 & $R_\sun$  &   Upper Uncertainty in $R_\star$  \\
28 & $M_\sun$ &   Stellar Mass ($M_\star$) \\
29 & $M_\sun$ &   Lower Uncertainty in $M_\star$  \\
30 & $M_\sun$ &   Upper Uncertainty in $M_\star$  \\
31 & K       &   Stellar Effective Temperature ($T_\textrm{eff}$)  \\
32 & K       &   Uncertainty $T_\textrm{eff}$  \\
33 & dex     &   Stellar Surface Gravity (log(g))  \\
34 & dex     &   Uncertainty log(g)  \\
35 & dex     &   Stellar Metallicity [Fe/H]  \\
36 & dex     &   Uncertainty [Fe/H]  \\
37 & ---     &   Stellar Spectral Classification \\
38 & ---     &   Visual Inspection Classification \\
\enddata
\tablecomments{This table is available in its entirety in machine-readable form.}
\end{deluxetable*}

\begin{figure}
\centering \includegraphics[width=\columnwidth{}]{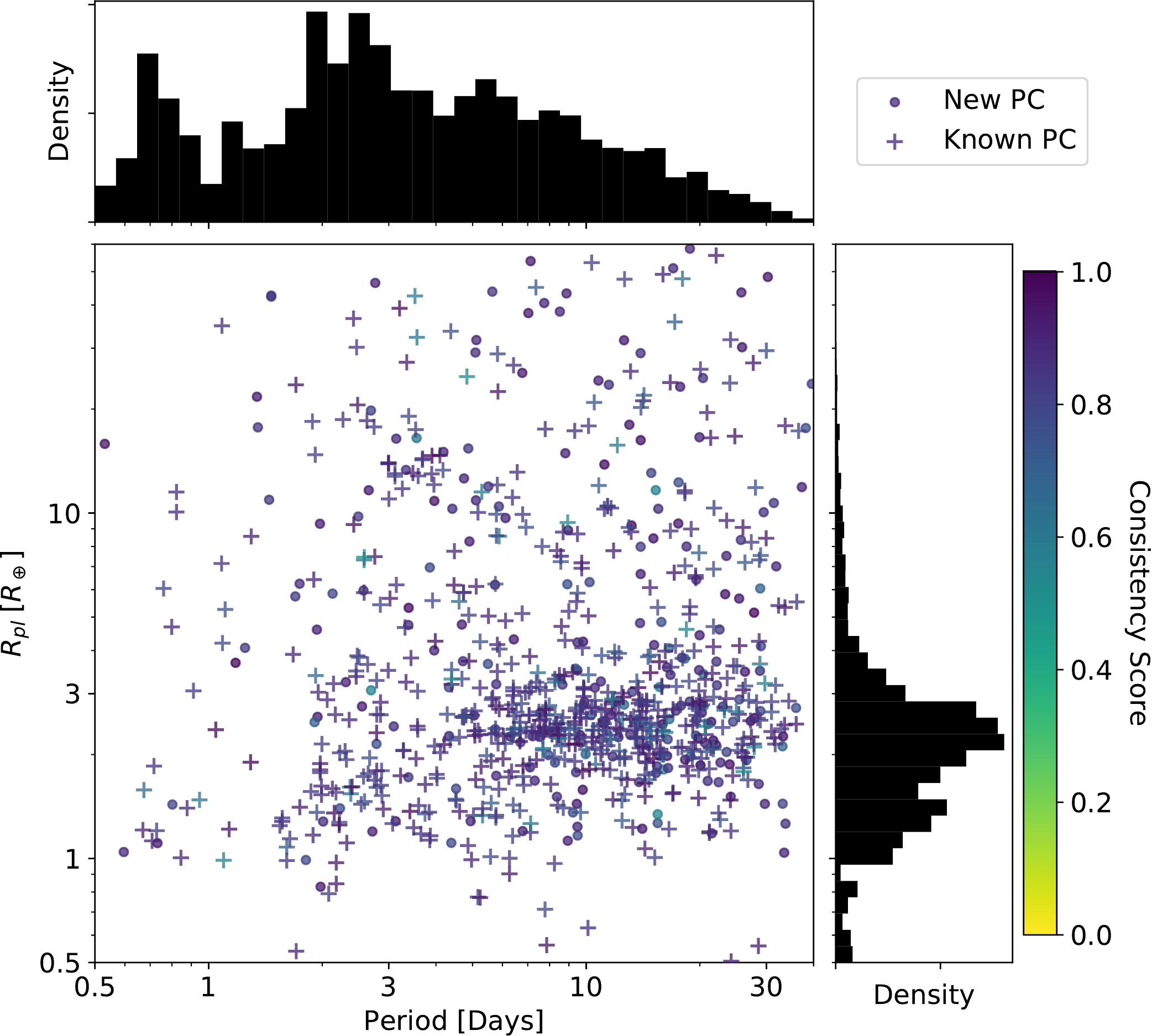}
\caption{The planet sample detected through our fully automated pipeline. The round markers show the new planet candidates (PCs) and + markers show the previously known PCs. The new candidates are uniformly distributed throughout the plot, because many of them come from campaigns not previously examined. The markers have been colored by consistency score to show that most candidates consistently passed the vetting metrics. 
\label{fig:catPl}}
\end{figure}

We provide our list of planetary parameters for this catalog with corresponding measurements of uncertainty in Table \ref{tab:catalog}. A plot of the resulting planet period and radius distribution is provided in Figure \ref{fig:catPl}. The deficit of planets near $2R_\earth$ is in alignment with the radius gap identified with \emph{Kepler} data \citep{ful17} and with previously discovered \emph{K2} candidates \citep{har20}. This gap appears to be indicative of some planetary formation or evolution mechanism, the exact origin of which remains unclear. One theory suggests stellar photoevaporation removes the envelope of weakly bound atmospheres in this region of parameter space \citep{owe17}, separating the super-Earths from the sub-Neptunes. Alternatively, the hot cores of young planets may retain enough energy to expel the atmosphere for planets within this gap (core-powered mass-loss; \citealt{gup18}). However, these two mechanisms require additional demographic data to parse out the main source of this valley. The catalog derived here provides additional planets and the necessary sample homogeneity needed to examine this feature with greater detail, but such a task is beyond the scope of the current work.

\subsection{Multi-Planet Yield}
\label{sec:mult}

\begin{figure}
\centering \includegraphics[width=\columnwidth{}]{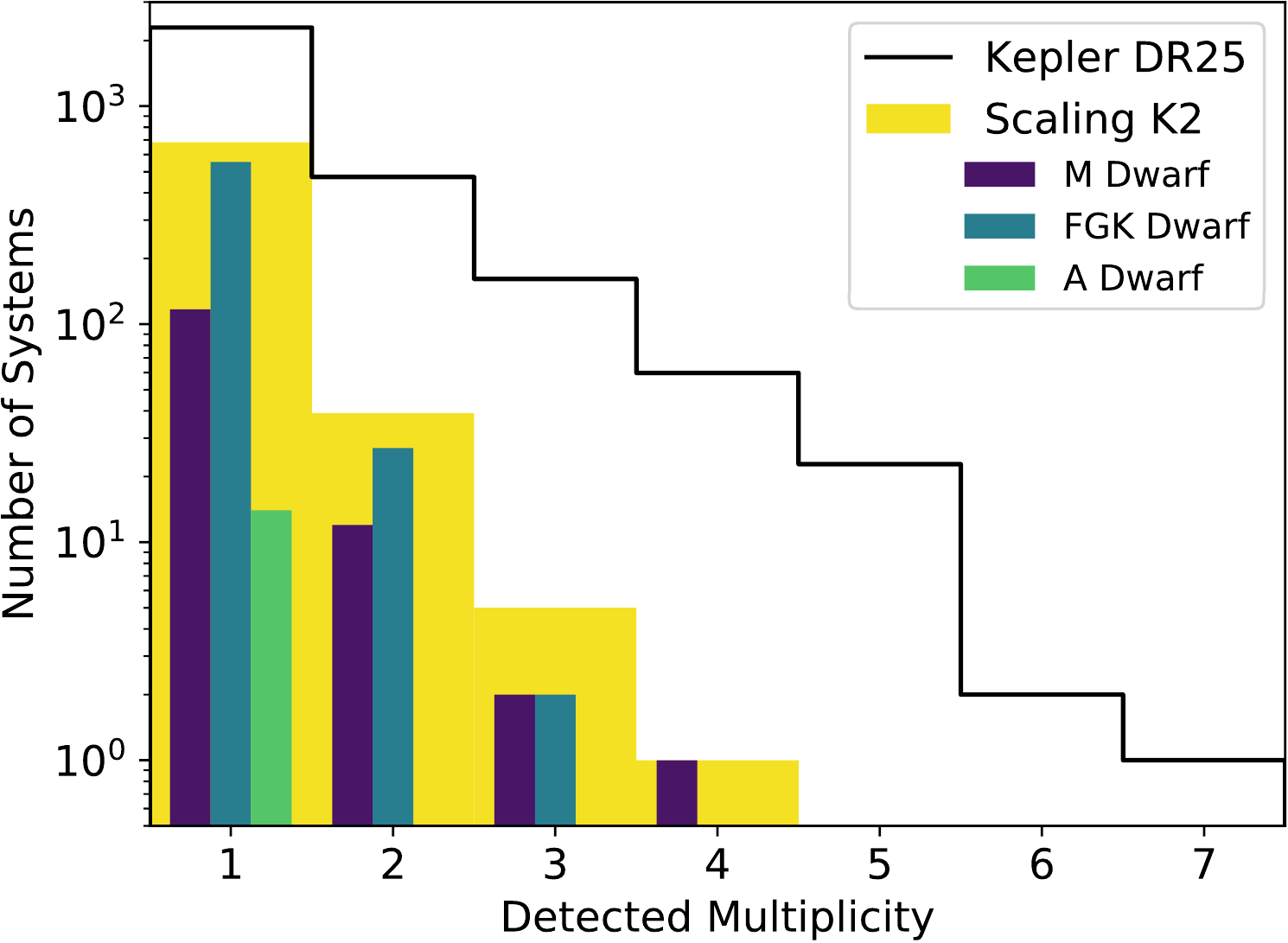}
\caption{ A histogram showing the observed system multiplicity distributions of our catalog (Scaling \emph{K2}) and the \emph{Kepler} DR25 \citep{tho18}. The longer data span and reduced noise properties enabled \emph{Kepler} to identify higher multiplicity systems. In addition, we provide the distributions as a function of stellar spectral type (M: $T_{\text{eff}}<4000K$; FGK: $4000\le T_{\text{eff}}\le 6500K$; A: $T_{\text{eff}}>6500K$).    
\label{fig:multHist}}
\end{figure}

Multi-planet systems provide a unique opportunity to understand the underlying system architecture and test intra-system formation mechanisms (e.g., \citealt{owe19}). Furthermore, these planets are more reliable, due to the unlikely probability of identifying two false positives in a light curve \citep{lis14, sin16}. In our sample we detected 57 unique multi-planet systems, with three systems independently identified in more than one campaign (EPIC 211428897, 212012119, 212072539). In Figure \ref{fig:multHist} we show the total observed multiplicity distribution for our catalog. Within this sample we did not find any multi-planet systems around A stars, but we identify 17 M dwarf systems and 40 FGK dwarf systems. In consideration of galactic latitude, we found 21 systems that lie greater than $40\degr$ above/below the galactic plane. EPIC 206135682, 206209135, and 248545986 all host three planet systems, while the remaining 18 systems only host two. This multi-planet sample provides unique coverage of galactic substructure.

\subsubsection{Unique Systems}

Our highest multiplicity system is EPIC 211428897, with four Earth-sized planets orbiting an M dwarf (system first identified by \citealt{dre17}). Our pipeline found this system independently in two of the overlapping fields (C5 and C18), further strengthening its validity. In addition, \citet{kru19} identified a fifth candidate with a period of 3.28 days. Despite the clear abundance of planet candidates, the \emph{K2} pixels span $3.98\arcsec$ and \emph{Gaia} DR2, which {\tt EDI-Vetter} uses to identify flux contamination, can only resolve binaries down to $1\arcsec$ for $\Delta \mathrm{mag}\lesssim3$ \citep{fur17}. Thus, high-resolution imaging is necessary for system validation. \citet{dre17} observed this system using Keck NIRC2 and Gemini DSSI, and found a companion star within $0.5\arcsec$. Since these planets are small, the likely $\sim \sqrt{2}$ radii increase, due to flux contamination, will not invalidate their candidacy \citep{cia17,fur17}. However, it remains unclear whether all of the planets are orbiting one of the stars or some mixture of the two. This will require further follow-up, and remains the subject of future work.

Through our analysis we identified 18 new multi-planet systems. EPIC 211502222 had been previously identified as hosting a single sub-Neptune with a period of 22.99 days by \citet{ye18}. Our pipeline discovered an additional super-Earth at 9.40 days, promoting this G dwarf to a multi-planet host. The remaining 17 systems are entirely new planet candidate discoveries, existing in campaigns not exhaustively searched (C12-C18). Notably, the K dwarf EPIC 249559552 hosts two sub-Neptunes which appear in a 5:2 mean-motion resonance. These resonant systems are important because of the potential detection of additional planets through transit-timing variations (e.g. \citealt{hol16}). EPIC 249731291 is also interesting since it is an early-type F dwarf (or sub-giant) system, with two short period gas giants.

\subsection{Low-Metallicity Planet Host Stars}
The core-accretion model indicates a link between stellar metallicity and planet formation \citep{pol96}. Observational evidence of this connection was first identified with gas giants \citep{san04, fis05}. However, the recent findings of \citet{tes19} complicates this narrative, by showing a lack of correlation between planetary residual metallicity and stellar metallicity. Additionally, direct comparison of metal-rich and metal-poor planet hosts in the \emph{Kepler} super-Earth and sub-Neptune populations, found no clear difference in planet occurrence \citep{pet18}. This is further indicative of a complex formation process. Thus, identification of low-metallicity planet hosts enables us to test our understanding of formation theories and to identify subtle features. Within our sample of planets, we found four candidates orbiting host stars with abnormally low stellar metallicity ([Fe/H]$<-0.75$). Upon further examination, two of these candidates (EPIC 212844216.01 and 220299658.01) have radii greater than $30R_\Earth$, indicative of an astrophysical false positive. The remaining two super-Earth sized planets orbit a low stellar metallicity M dwarf (EPIC 210897587; [Fe/H]$=-0.831\pm0.051$). This system has previously been validated using WIYN/NESSI high resolution speckle imaging \citep{hir18} and appears in tension with the expectations of core-accretion in-situ formation, providing constraints for planet formation models.

\section{Measuring Completeness}
\label{sec:complete}

Any catalog of planets will inherit some selection effects due to the methodology of detection, limitations of the instrument, and stellar noise. These biases will affect the sample completeness and must be accounted for when conducting a demographic analysis. The selection of transiting planets can be addressed using analytic arguments, but the instrument and stellar noise contributions to the sample completeness are dependent on the stellar sample and the specifics of the instrument. With an automated detection pipeline this detection efficiency mapping can be achieved through the implementation of an injection/recovery test. Here, artificial signals are injected into the raw photometry and run through the automated software to test the pipeline's recovery capabilities, directly measuring the impact of instrument and stellar noise on the catalog. Many previous studies have used this technique on \emph{Kepler} data, yielding meaningful completeness measurements \citep{pet13b,chr13,chr15,chr20,bur17}.

\subsection{Measuring CDPP}
\label{sec:cdpp}
The first step in computing the detection efficiency is to establish a noise profile for each target light curve. With these values available, the signal strength can be estimated given the transit parameters. The ability to make these calculations enables one to understand the likelihood of detection for a given signal. \emph{Kepler} used the combined differential photometric precision (CDPP) metric to quantify the expected target stellar variability and systematic noise, as described in \citet{chr12}. For all targets we provide CDPP measurements for following transit durations: 1, 1.5, 2, 2.5, 3, 4, 5, 6, 7, 8, 9, and 10 hours, spanning the range of occultation timescales expected for \emph{K2} planet candidates.

To compute these values we randomly injected a weak ($3\sigma$) transit signal, with the appropriate transit duration, into each detrended light curve. The strength of the recovered signal, normalized by the injected signal depth, provided a measure of the target CDPP. This process was carried out 450 times to ensure a thorough and robust examination of the light curve, sampling the full light curve for four hour transits ($\sim$20 day period) and 25\% of the light curve for one hour transits ($\sim$0.5 day period). To measure the impact of differing transit durations, this process was executed for each respective CDPP timescale. For a detailed account of this procedure see Section 4 of \citet{zin20a}. 

\begin{figure}
\centering \includegraphics[width=\columnwidth{}]{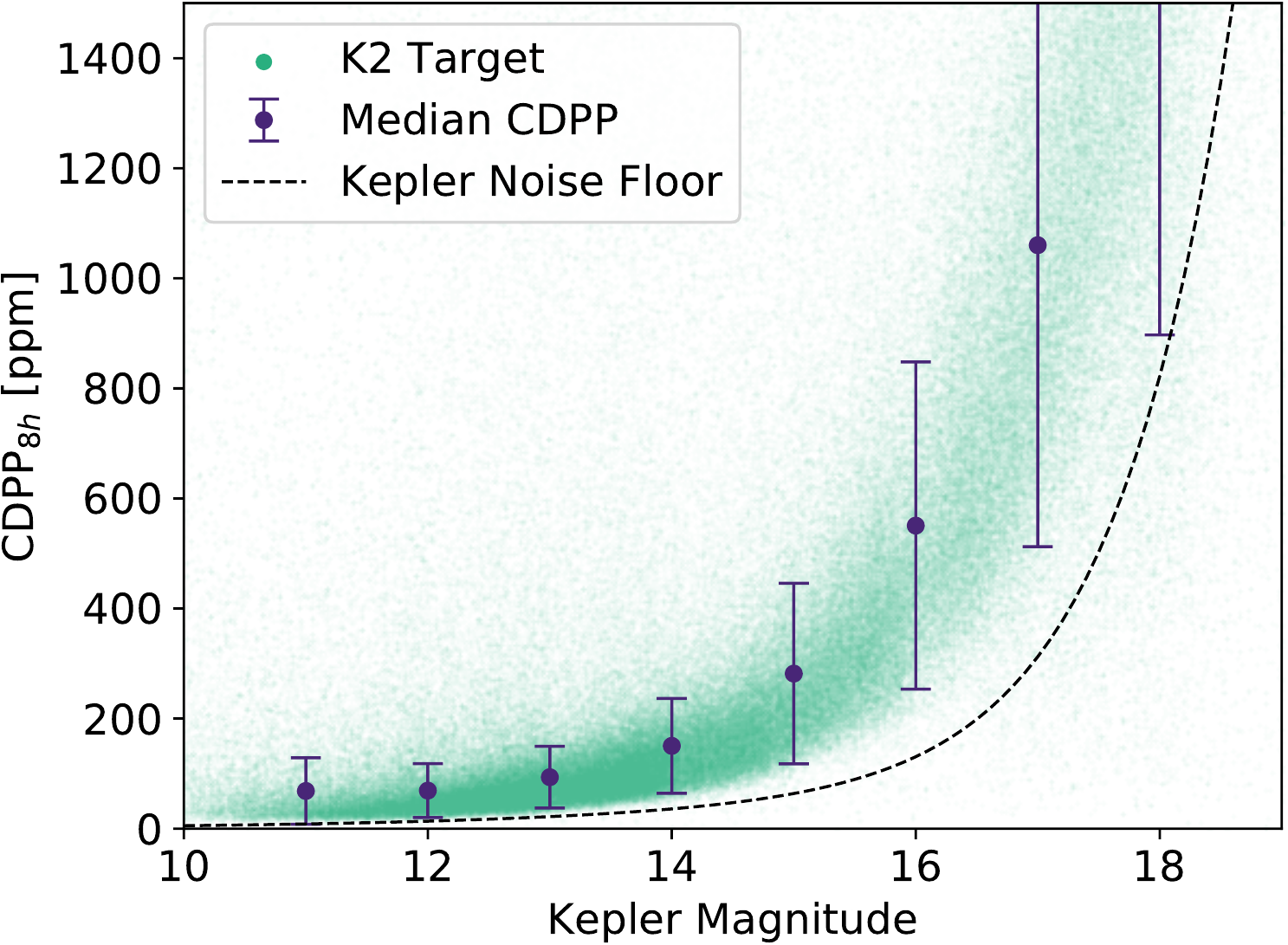}
\caption{The light curve noise (the 8-hour CDPP measurements) for our target sample as a function of the \emph{Kepler} broadband magnitude measurement. The median CDPP markers represent the median within a one magnitude bin and the corresponding bin standard deviation. The \emph{Kepler} noise floor represents the shot and read noise expected from the detector alone \citep{jen10b}.
\label{fig:cdpp}}
\end{figure}

In Figure \ref{fig:cdpp} we show the measured 8h CDPP for the targets within this sample.  Overall, there is a clear correlation with photometric magnitude, demonstrating our ability to quantify the light curve noise properties. Moreover, despite the introduction of additional systematic noise from the spacecraft pointing, the detrended \emph{K2} photometry is still near the theoretical noise limit at the brighter magnitudes. The complete set of measurements are available in Table \ref{tab:cdpp}.

\begin{deluxetable*}{lcc}
\tablecaption{Description of the CDPP measurements of each stellar target. \label{tab:cdpp}}
\tablehead{\colhead{Column} & \colhead{Units} & \colhead{Explanation}} 
\startdata
1 & ... & EPIC identifier\\
2 & ... & Campaign\\
3 & ppm & CDPP RMS Value for Transit of 1.0 hr \\
4 & ppm & CDPP RMS Value for Transit of 1.5 hr \\
5 & ppm & CDPP RMS Value for Transit of 2.0 hr \\
6 & ppm & CDPP RMS Value for Transit of 2.5 hr \\
7 & ppm & CDPP RMS Value for Transit of 3.0 hr \\
8 & ppm & CDPP RMS Value for Transit of 4.0 hr \\
9 & ppm & CDPP RMS Value for Transit of 5.0 hr \\
10 & ppm & CDPP RMS Value for Transit of 6.0 hr \\
11 & ppm & CDPP RMS Value for Transit of 7.0 hr \\
12 & ppm & CDPP RMS Value for Transit of 8.0 hr \\
13 & ppm & CDPP RMS Value for Transit of 9.0 hr \\
14 & ppm & CDPP RMS Value for Transit of 10.0 hr \\
\enddata
\tablecomments{This table is available in its entirety in machine-readable form.}
\end{deluxetable*}

\subsection{Injection/Recovery}

There are several points along the pipeline at which the signal can be injected. Ideally, injections would be made on the rawest form of photometry (at the pixel-level), but doing so is computationally expensive and provides a marginal gain in completeness accuracy (see \citealt{chr17} for the effects on the \emph{Kepler} data set). Moving just one step downstream, the injections can be more easily made at the light curve-level. Here, the artificial signal is introduced into the aperture-integrated flux measurements, followed by pre-processing, detrending and signal detection. Finally, the most accessible, but least accurate, method is by injecting signals after pre-processing (e.g., \citealt{clo20}). Since the pre-processed {\tt EVEREST} light curves are readily available, this method requires minimal computational overhead. However, it fails to capture the impact of pre-processing on the sample completeness. These effects are especially important for \emph{K2} photometry, which undergoes significant modification before being searched. Following the procedures of our previous study \citep{zin20a}, we injected our artificial signals into the aperture-integrated light curves (before pre-processing; see Figure \ref{fig:diagram}). In the next few paragraphs we briefly outline our methodology, but suggest interested readers reference Section 5 of our previous work for a more detailed account.

Using the {\tt batman} Python package \citep{kre15}, we created and injected artificial transits in the raw flux data. For each target, we uniformly drew a period from [0.5, 40] days and an $R_{pl}/R_{\star}$ from a log-uniform distribution with a range [0.01,0.1]. The ephemeris was uniformly selected, with the requirement that at least three transits reside within the span of the light curve, and the impact parameter was uniformly drawn from [0,1].\footnote{In the current iteration of this pipeline we have removed the eclipsing binary impact parameter limit mentioned in Section 5 of \citet{zin20a}. In doing so, we provide a more accurate accounting of the impact of grazing transits.} All injections were assumed to have zero eccentricity. This assumption is motivated by the short period range of detectable \emph{K2} planets, which many have likely undergone tidal circularization. In addition, eccentricity only affects the transit duration, making its impact on completeness minor. The limb-darkening parameters for the artificial transits were dictated by the stellar parameters discussed in Section \ref{sec:stellarSample}. Using the ATLAS model coefficients for the \emph{Kepler} bandpasses \citep{cla12}, we derived the corresponding quadratic limb-darkening parameters based on their stellar attributes. In cases where stellar parameters did not exist, we assumed solar values.

We expect the pipeline's recovery capabilities scale as a function of the signal strength. In order to quantify this effect, one must have a measure of the expected injection signal strength (MES). Equipped with our CDPP measurements, this value is directly related to the transit depth (depth) and can be analytically found using:
\begin{equation}
\text{MES}=C\;\frac{\text{depth}}{\text{CDPP}_{\text{t}_{\text{dur}}}}\;\sqrt{N_{\text{tr}}},
\label{eq:MES}
\end{equation}
where $\text{CDPP}_{\text{t}_{\text{dur}}}$ represents the targets CDPP measures for a given transit duration (achieved through interpolation of the measured CDPP values discussed in Section \ref{sec:cdpp}) and $N_{\text{tr}}$ is the number of available transits within the data span. The $C$ value is a global correction factor that renormalizes the analytic equation to match the detected signal values. For this data set we found $C=0.9488$. Using this equation, we calculated the expected MES values for all injections and measured the sample completeness.

\begin{figure*}
\centering \includegraphics[width=\textwidth{}]{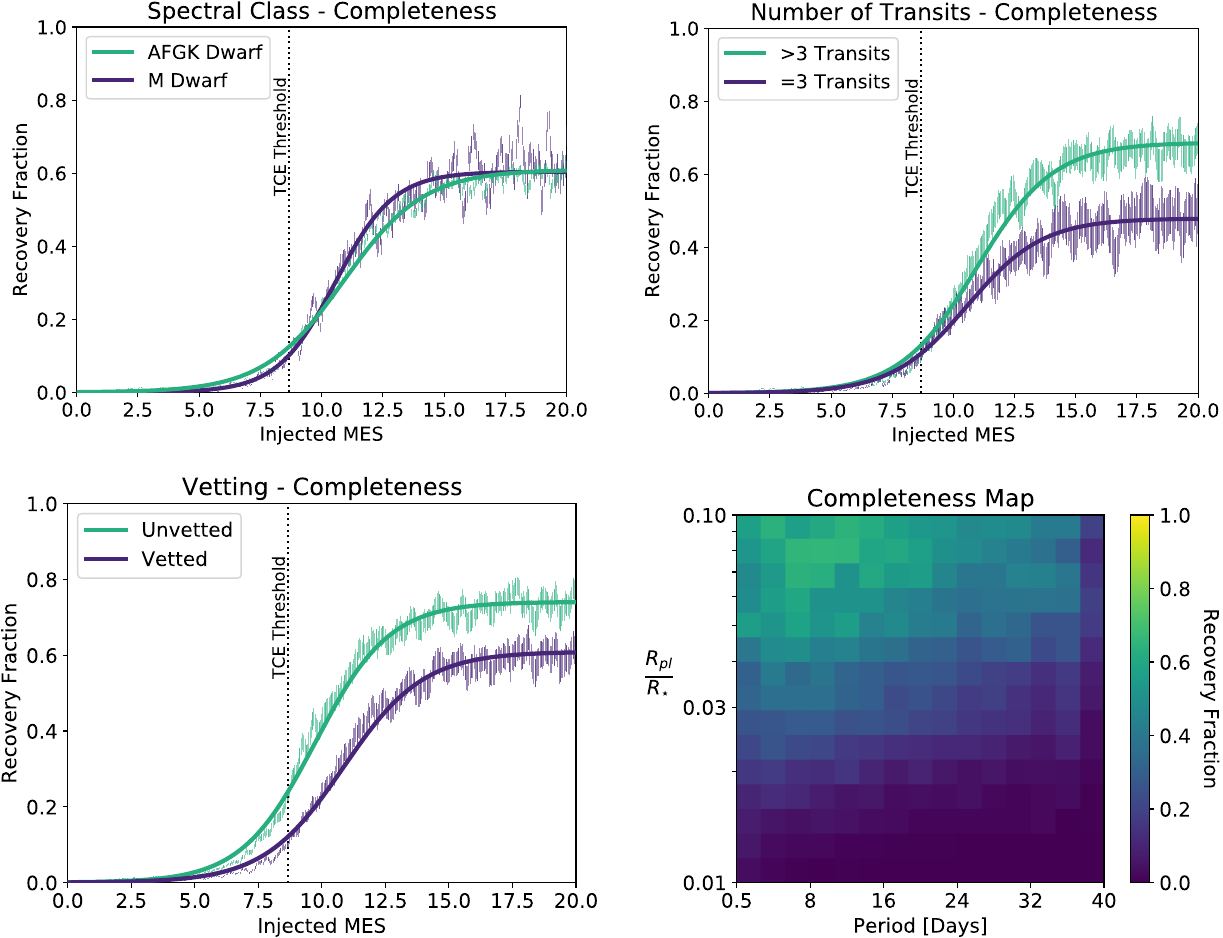}
\caption{Plots of the measured completeness, resulting from our injection/recovery test, for the presented planet sample. To show the effects of characteristic stellar noise, the number of transits, and the loss of planets due to our vetting software, we provide several slices of the data. The corresponding logistic function parameters are available in Table \ref{tab:complete}. The heat map shows the overall vetted completeness as a function of planet radius ratio and the transit period. For our stellar sample, the radii ratios of 0.01, 0.03, and 0.1 correspond to median planet radii of 1.1, 3.4, and 11.3 $R_\Earth$ respectively.  
\label{fig:TCEcomplete}}
\end{figure*}

Once our injections were performed, we passed this altered photometry through the software pipeline, testing our recovery capabilities. We considered a planet successfully recovered if it met the following criteria: the detected signal period and ephemeris were within $3\sigma$ of the injected values and the signal passes all of the vetting metrics. The results of this test can be seen in the completeness map in Figure \ref{fig:TCEcomplete}. To quantify our detection efficiency as a function of injected MES, we used a uniform kernel density estimator (KDE; width of 0.25 MES) to measure our software's recovery fraction. These values were then fit with a logistic function of the form:
\begin{equation}
f(x)=\frac{a}{1+e^{-k(x-l)}}.
\end{equation}
The best fit values of $a$, $k$, and $l$ are listed in Table \ref{tab:complete}.

While MES is closely related to completeness, additional signal parameters can play a role. \citet{chr20} tested the effects of stellar effective temperature ($T_{\text{eff}}$), period, $N_{\text{tr}}$, and photometric magnitude on the completeness of the analogous \emph{Kepler} injections, finding the strongest effect linked to $N_{\text{tr}}$. In Figure \ref{fig:TCEcomplete} we show two of these completeness features for the \emph{K2} injections: spectral class and $N_{\text{tr}}$. \citet{chr15} noted a significant drop ($\sim4\%$) in detection efficiency for cooler M dwarfs, which exhibit higher stellar variability. Remarkably, we did not find a significant difference between the AFGK dwarfs ($T_{\text{eff}}>4000K$) and the M dwarfs ($T_{\text{eff}}<4000K$). It is likely that the increased systematic noise of \emph{K2} blurs this completeness feature, making the two populations indistinguishable. We also considered the additional noise contributions expected for young stars. In looking at 8033 targets associated with young star clusters, as indicated by {\tt BANYAN $\Sigma$} \citep{gag18}, we could not identify a significant completeness difference. Like the \emph{Kepler} TPS, we found $N_{\text{tr}}$ has the strongest effect on our \emph{K2} planet sample completeness. Here, a significant drop in detection efficiency was expected for signals near the minimum transit threshold. In light of the pipeline's three transit requirement, these marginal signals were more susceptible to systematics and vetting misclassification. In other words, if any of the three transits were discarded by the vetting, it would immediately receive a false positive label. We also parsed the data into larger $N_{\text{tr}}$ value bins, but found little difference in completeness between $N_{\text{tr}}$ equal to four, five, and six; the vetting was unlikely to discard more than one meaningful transit. Furthermore, we considered the effects of the signal period. While this parameter is strongly correlated with $N_{\text{tr}}$, it has the potential to describe period-dependent detrending and data processing issues. Separating the injected signals at a period of 26 days (see Table \ref{tab:complete}), we found a loss of completeness ($\Delta a\sim0.20$) that was comparable to the $N_{\text{tr}}$ partition ($\Delta a\sim0.21$). This similarity indicates a strong correlation between parameters, suggesting either of the two features would appropriately account for the reduced completeness in this region of parameter space. Since $N_{\text{tr}}$ provides a slightly larger deficit, we suggest using this function for future demographic analysis of our catalog.  

Completeness measurements provide a natural test of {\tt EDI-Vetter}'s classification capabilities. The introduction of systematics from the telescope makes signal vetting more difficult, when compared to the \emph{Kepler} TPS, requiring more drastic vetting metrics. This could lead to significant misclassification, discarding an abundance of meaningful planet candidates. Fortunately, in Figure \ref{fig:TCEcomplete} we identify a $\sim13\%$ loss of completeness due to the vetting metrics, which is comparable to the \emph{Kepler} {\tt Robovetter} ($\sim10\%$ loss; \citealt{cou17b}). Ideally, this difference would be zero, but such a minimal loss is acceptable and, more importantly, quantifiable. 

We have provided the completeness parameters necessary for a demographic analysis of this catalog. However, the injections carried out here are computationally expensive and hold significant value for research beyond the scope of this catalog. Therefore, we provide, in addition, the injected {\tt EVEREST}-processed light curves and a summary table of the injection/recovery test.\footnote{\href{http://www.jonzink.com/scalingk2.html}{http://www.jonzink.com/scalingk2.html}} These data products enable users to set custom completeness limits and to test their own vetting software with significantly reduced overhead.

\begin{deluxetable}{lccc}
\tablecaption{The logistic parameters for the corresponding completeness functions shown in Figures \ref{fig:TCEcomplete} \& \ref{fig:complete_lat}. Additionally, we include the completeness parameters for a separation of 26 day period signals. It is important to highlight that $a$ represents the maximum completeness for high MES signals. \label{tab:complete}}
\tablehead{\colhead{Model} & \colhead{a} & \colhead{k} & \colhead{l}} 
\startdata
\textbf{Unvetted} & 0.7407 & 0.6859 & 9.7407 \\
\textbf{Vetted} & 0.6093 & 0.6369 & 10.8531 \\
\\
\textbf{>3 Transits} & 0.6868 & 0.6347 & 10.9473\\
\textbf{=3 Transits} & 0.4788 & 0.6497 & 10.5598\\
\\
\textbf{<26d Periods} & 0.6619 & 0.6231 & 10.9072\\
\textbf{>26d Periods} & 0.4635 & 0.6607 & 10.5441\\
\\
\textbf{AFGK Dwarf} & 0.6095 & 0.6088 & 10.8986\\
\textbf{M Dwarf} & 0.6039 & 0.8455 & 10.5636 \\
\\
\textbf{C1} & 0.3923 & 0.7654 & 11.3914\\
\textbf{C2} & 0.6430 & 0.7173 & 10.8544 \\
\textbf{C3} & 0.7462 & 0.6689 & 10.5701 \\
\textbf{C4} & 0.6734 & 0.6344 & 11.1443 \\
\textbf{C5} & 0.4425 & 0.5923 & 11.3923 \\
\textbf{C6} & 0.7654 & 0.5759 & 10.8772 \\
\textbf{C7} & 0.3941 & 0.6052 & 11.7002 \\
\textbf{C8} & 0.6669 & 0.5726 & 10.0560 \\
\textbf{C10} & 0.5572 & 0.6469 & 10.0056 \\
\textbf{C11} & 0.2171 & 0.4759 & 12.3882 \\
\textbf{C12} & 0.6192 & 0.7341 & 10.6272 \\
\textbf{C13} & 0.6853 & 0.5698 & 11.3878 \\
\textbf{C14} & 0.7505 & 0.6596 & 10.9776 \\
\textbf{C15} & 0.6067 & 0.6480 & 10.4673 \\
\textbf{C16} & 0.6809 & 0.7256 & 10.5453 \\
\textbf{C17} & 0.5848 & 0.6633 & 10.3635 \\
\textbf{C18} & 0.6116 & 0.4676 & 11.5783 \\
\enddata

\end{deluxetable}

\subsection{Completeness and Galactic Latitude}
\label{sec:comLat}
\begin{figure*}
\centering \includegraphics[width=\textwidth{}]{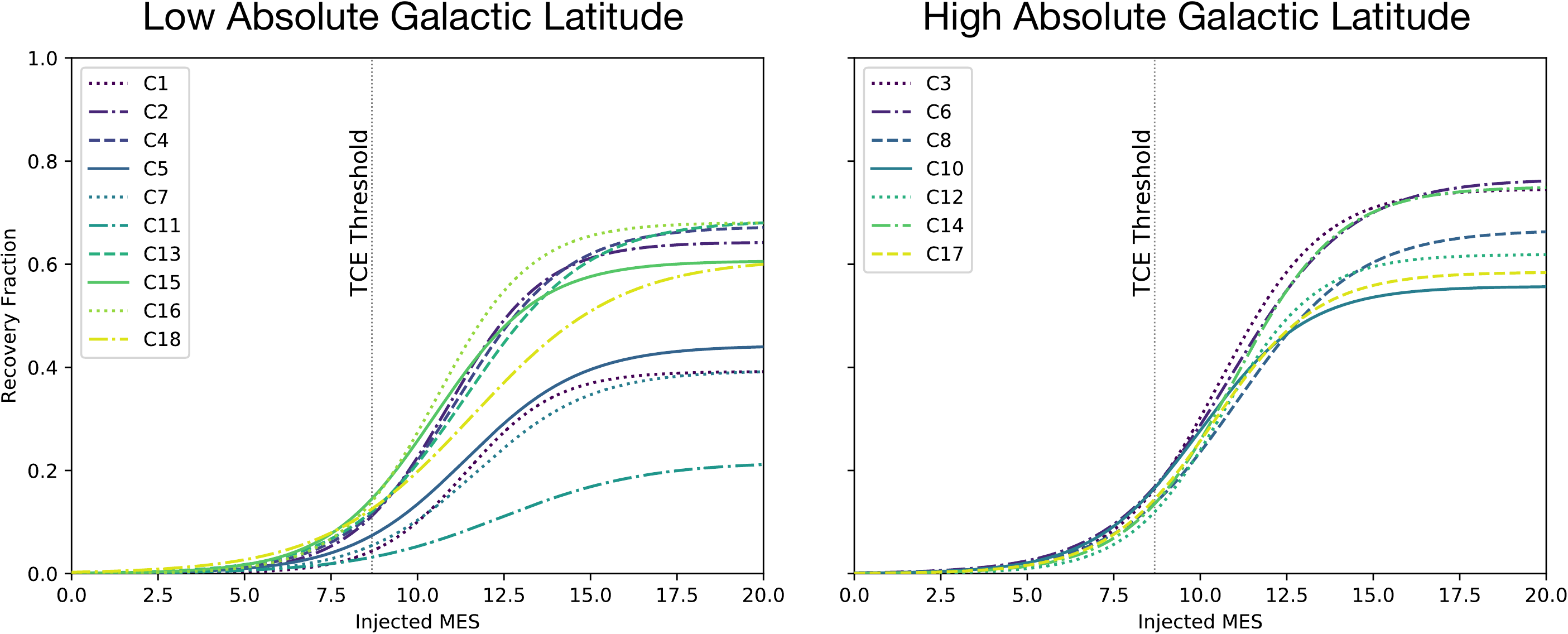}
\caption{The calculated vetted completeness for the low (left; $\mid b \mid<40\degr$) and high (right; $\mid b \mid>40\degr$) galactic latitude Campaigns. The corresponding logistic function parameters are available in Table \ref{tab:complete}. 
\label{fig:complete_lat}}
\end{figure*}

Each \emph{K2} campaign probed a different region along the ecliptic. These fields correspond to unique galactic latitudes, where distinct noise features may spawn differences in inter-campaign completeness. To address these potential variations, we consider the completeness as a function of campaign in Figure \ref{fig:complete_lat}.

Overall, there is significant scatter among the lower absolute galactic latitude campaigns ($\mid b \mid<40\degr$). This trend is indicative of target crowding near the galactic plane \citep{gil15}. In these low latitude fields the photometric apertures are more contaminated by background sources, contributing additional noise, variability, and flux dilution, making transit detection more onerous. This is highlighted by Campaign 11, which is the closest field to the galactic plane ($b\sim9\degr$) that we analyzed, thus providing the lowest completeness of all campaigns ($a=0.22$). Conversely, Campaign 6 with $b\sim48\degr$ has the highest completeness ($a=0.77$).

In the high galactic latitude campaigns ($\mid b \mid>40\degr$), this crowding effect is less salient, reducing the inter-campaign completeness scatter. Furthermore, these well isolated targets, provide high quality photometry and yield the highest completeness of all \emph{K2} campaigns ($a\sim 0.70$).

These inter-campaign differences are meaningful, however the limited number of targets (and synthetic transit injections) within each campaign, subjects these completeness measurements to further uncertainty. Therefore, the values provided for the global completeness assessment (for Figure \ref{fig:TCEcomplete}) are more robust and should be used for full catalog occurrence analysis.

\section{Measuring Reliability}
\label{sec:reli}
Despite efforts to remove problematic cadences, instrument systematics pollute the light curves, creating artificial dips that can be erroneously characterized as a transit signal.
To measure the reliability in a homogeneous catalog of planets, the rate of these false alarms (FAs) must be quantified. \citet{bry19} showed that proper accounting of the sample FA rate is essential in extracting meaningful and consistent planet occurrence measurements. 

The main goal of {\tt EDI-Vetter} (and its predecessor {\tt Robovetter}) is to parse through all TCEs and remove FAs without eliminating true planet candidates. However, this process is difficult to automate and requires a method of testing the software's capability to achieve this goal. Accomplishing such a task necessitates an equivalent data set that captures all the unique noise properties, which contributes to FAs, without the existence of any true astrophysical signals. With such data available, the light curves can be processed through the detection pipeline. If the vetting algorithm worked perfectly, nothing would be identified as a planet candidate. Therefore, any signals capable of achieving planet candidacy would be an authentic FA and provide insight into the software's capabilities. 

\citet{cou17b} explored two methods for simulating this necessary data using the existing light curves: data scrambling and light curve inversion. The first method takes large portions of the real light curve data and randomizes the order, retaining all of the noise properties while scrambling out true periodic planet signatures. Using this method \citep{tho18} was able to replicate the real long period TCE distribution of the \emph{Kepler} DR25 catalog. The second method inverts the light curve flux measurements. Upon this manipulation, the existing real transit signals will now be seen as flux brightening events, rendering them unidentifiable by the transit search algorithm. Under the assumption that many systematic issues are symmetric upon a flux inversion, the photometry will retain its noise properties without containing any real transit signals. This method was used by \citet{tho18} to replicate the real short period TCE distribution of the \emph{Kepler} DR25 catalog, while maintaining the quasi-sinusoidal features, like the rolling bands that repeat due to the spacecraft's temperature fluctuations (see Section 6.7.1 of \citet{van16} for further detail on this effect).

\begin{figure}
\centering \includegraphics[width=\columnwidth{}]{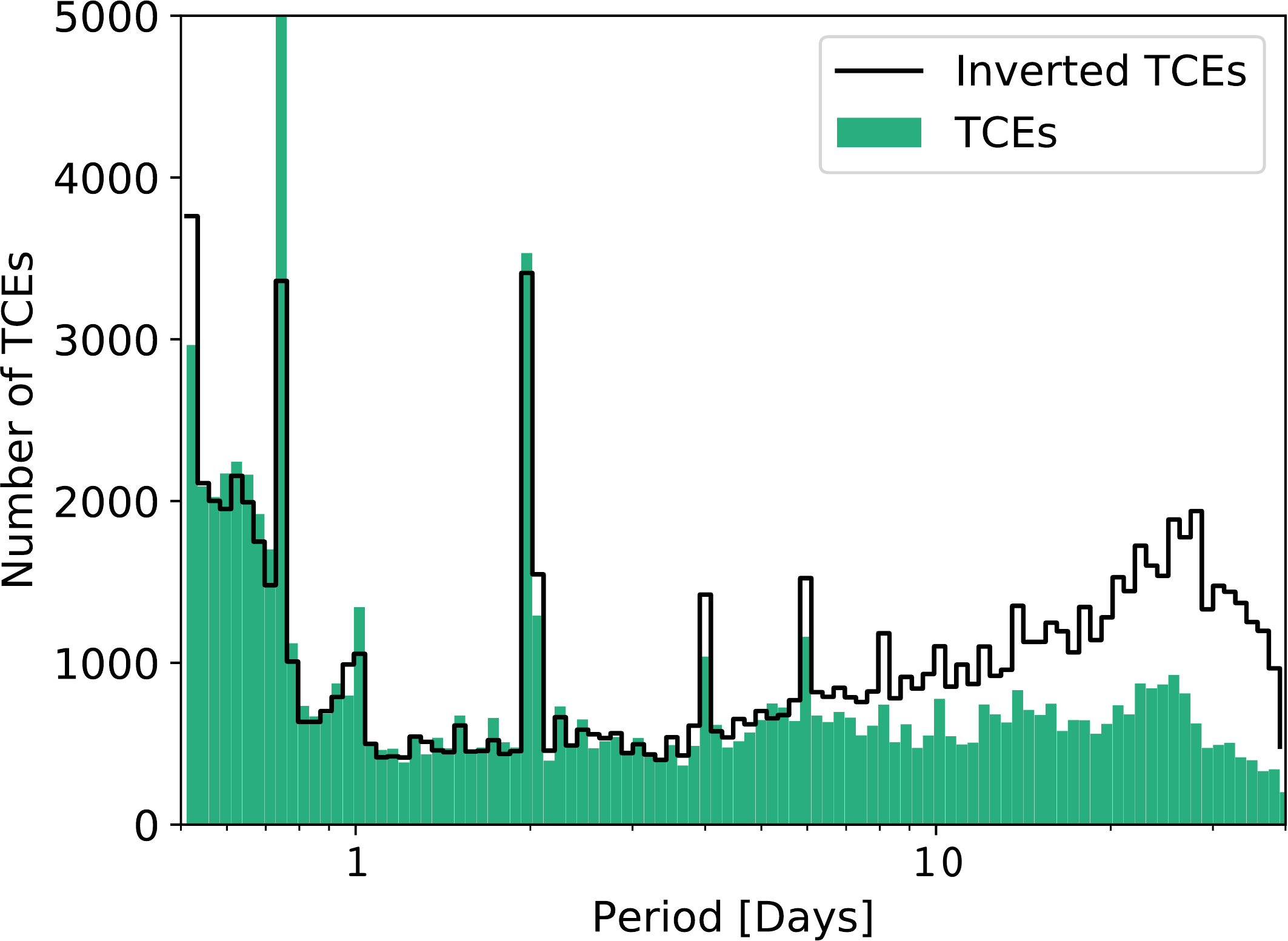}
\caption{The distribution of TCEs from the real light curves and the distribution of inverted TCEs from the FA simulation. This shows consistency among the two distributions with a minor surplus of inverted TCEs at longer periods and a deficiency at the 3rd harmonic of the thruster firing (18 hours).
\label{fig:TCEreliability}}
\end{figure}

Since the spacecraft also underwent quasi-periodic roll motion during the \emph{K2} mission, leading to cyclical instrument systematics, we chose the light curve inversion method. In Figure \ref{fig:TCEreliability} we assess our ability to capture the noise features using this method. By comparing the distribution of TCEs from the real light curves with those of the inverted light curves, we can identify regions of parameter space where the inversions over- or under-represent systematic noise properties. Overall, the two distributions (108,379 TCEs and 110,548 inverted TCEs)\footnote{The individual transit check described in Section \ref{sec:ITC} was the dominant vetting metric for both the TCEs and the inverted TCEs, discarding non-transit shaped signals} are well aligned, providing an adequate simulation of the data set's noise characteristics. However, the inverted TCEs are slightly over-represented at long periods and under-represented at the 3rd harmonic of the 6 hour thruster firing. While we acknowledge these minor discrepancies, their impact will be small on the overall measure of the catalog reliability. 

After running the inverted light curves through our pipeline we identify 77 FA signals as planet candidates ( 806 candidates were identified in the real light curves). We used this information, alongside the formalism described in Section 4.1 of \citet{tho18}, to quantify our sample reliability. The first step in achieving this measurement requires an understanding of the vetting routines FA removal efficiency ($E$). From the inverted light curve test we can estimate $E$ as
\begin{equation}
E\approx\frac{N_{FP_{\mathrm{inv}}}}{T_{TCE_{\mathrm{inv}}}},
\end{equation}
where $N_{FP_{\mathrm{inv}}}$ is the number of TCEs that were accurately flagged as false positives and $T_{TCE_{\mathrm{inv}}}$ is the total number of TCEs found in the inversion test. For the total data set, we found {\tt EDI-Vetter} has a 99.9\% efficiency in removing FA signals. However, this extreme competence must be balanced by the abundance of TCEs found by our pipeline. 

\begin{figure}
\centering \includegraphics[width=\columnwidth{}]{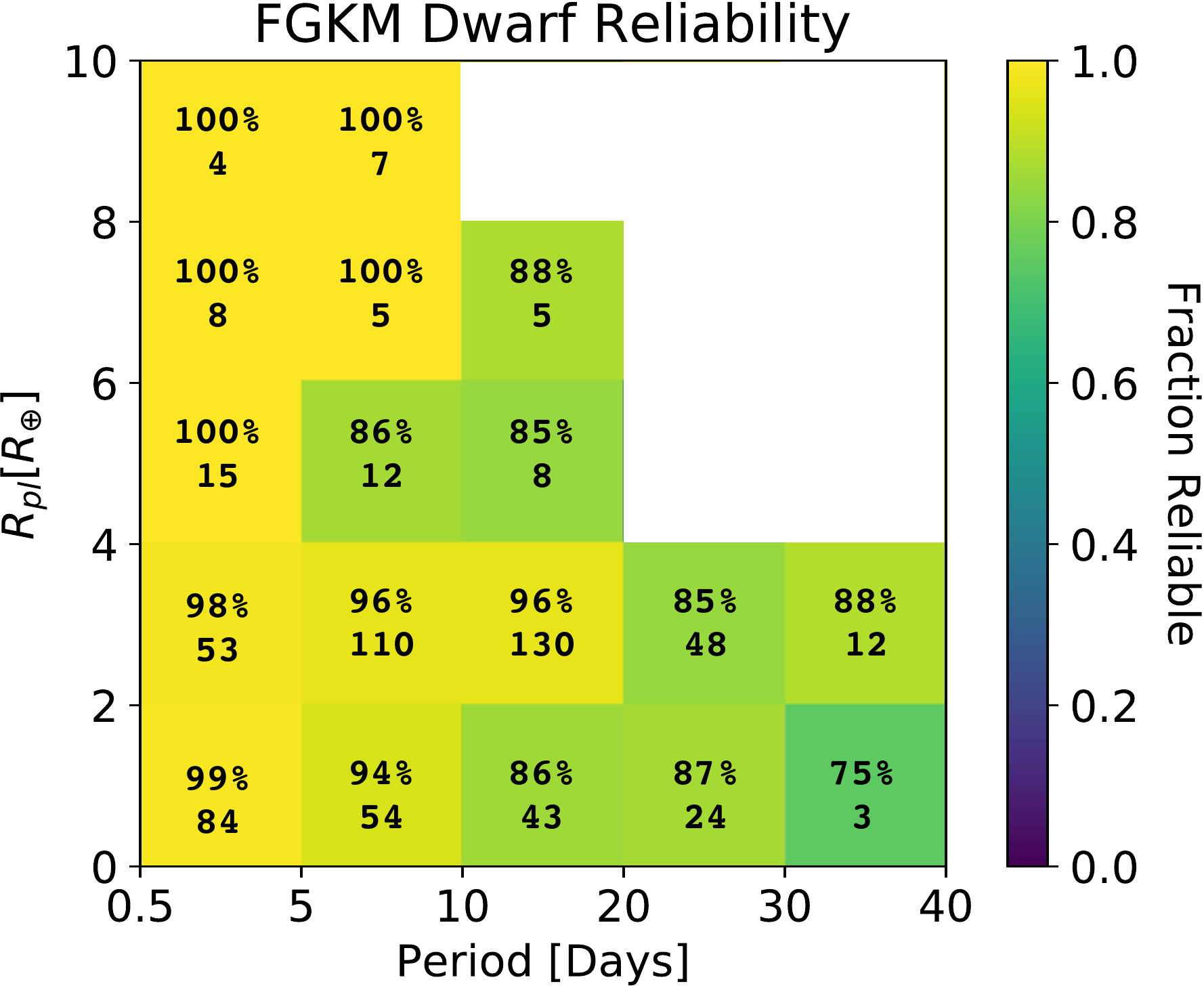}
\caption{The calculated reliability of our planet sample as a function of orbital period and planet radius. The reliability percent and the number of candidates have been listed in each corresponding box. The white regions represent areas of parameter space where the number of candidates and FAs are sparse, making accurate measurements of reliability unachievable. 
\label{fig:reliability}}
\end{figure}

We can determine the reliability fraction ($R$) of our catalog using the number of TCEs flagged as false positive ($N_{FP}$) in the real light curves and the number of planet candidates ($N_{PC}$):
\begin{equation}
R=1-\frac{N_{FP}}{N_{PC}}\Bigg(\frac{1-E}{E}\Bigg).
\label{eq:reli}
\end{equation}
Overall, we found the planet catalog provided here is 91\% reliable. This is slightly lower than the 97\% reliability of the \emph{Kepler} DR25 catalog \citep{tho18}. However, this \emph{Kepler} value is greatly improved by the extensive data baseline. A more appropriate comparison would consider a period range with a comparable number of transits (\emph{Kepler} candidates with periods greater than 10 days have a similar number of transits). In this region of parameter space the {\tt Robovetter} has a reliability of 95\%, which is closer to the value reported for {\tt EDI-Vetter}. Moreover, these broad summary statistics fail to capture the complexity of this metric.

\begin{figure*}
\centering \includegraphics[height=7cm]{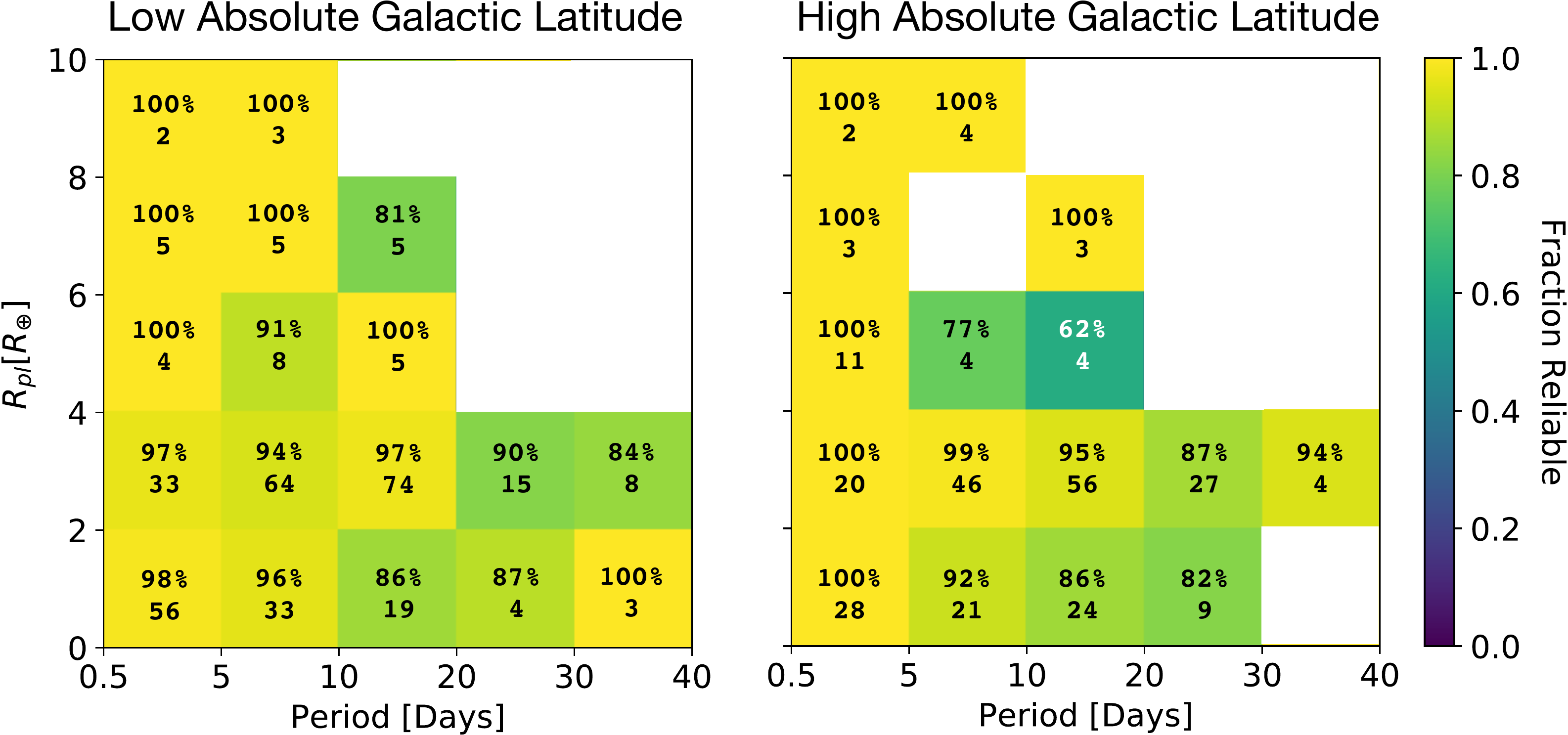}
\caption{The calculated reliability for the low and high galactic latitude Campaigns. The reliability percent and the number of candidates have been listed in each corresponding box. The white regions represent areas of parameter space where the number of candidates and FAs are sparse, making accurate measurements of reliability unachievable. 
\label{fig:reliability_lat}}
\end{figure*}

We found 8 of the FAs detected in our inversion simulation are hosted by sub-giant and giant stars, which have notably more active stellar surfaces. By excluding all giants with $\log(g)$ less than 4, we focus on the dwarf star population (in Figure \ref{fig:reliability}) and consider how reliability changes as a function of period and planet radius. At longer periods the reliability drops. The reduced number of transits available for a given candidate more easily enables systematic features to line up and create a FA signal. Smaller planet radius regions are also more susceptible to FAs due to their weak signal strength, which can be replicated by noise within the data set. When accounting for this period and radius dependence we expect 94\% of our dwarf host candidates to be real astrophysical signals.

\subsection{Reliability and Galactic Latitude}
\label{sec:galRelib}

Like completeness, reliability may also exhibit inter-campaign differences. However, the number of FAs detected by inversion are not significant enough to enable thorough campaign by campaign analysis. Since Section \ref{sec:comLat} showed that a majority of field differences could be attributed to galactic latitude, we considered differences in reliability for high ($\mid b \mid>40\degr$) and low ($\mid b \mid<40\degr$) absolute galactic latitudes in Figure \ref{fig:reliability_lat}. Overall, we found that both low and high absolute galactic latitude campaigns produce a very similar reliability, 95\% and 94\% respectively. Upon examination of these two stellar populations, no significant differences were identified, thus these changes in reliability are likely due to statistical fluctuations from the reduced number of FAs (34 in low latitudes and 35 in high latitudes). Therefore, we encourage future population analysis to use the full reliability calculation provided in Figure \ref{fig:reliability}.

\subsection{Astrophysical False Positives}
\label{sec:astrFP}
It is important to highlight that reliability is a measure of the systematic FA contamination rate. There exist non-planetary astrophysical sources capable of producing a transit signal. For example, dim background eclipsing binaries (EBs) may experience significant flux dilution from the primary target, manifesting a shallow depth transit. This planetary signal mimicry may lead to candidate misclassification \citep{fre13,san16,mor16,mat18}.  The \emph{Kepler} DR25 relied on the centroid offset test \citep{mul17} to identify these contaminants. This test considered the TCE's flux difference in and out of transit for each aperture pixel. Larger differences near the edge (or away from the center) of the aperture indicated a non-target star origin, enabling the identification of contaminants to within $1\arcsec$. Although it is possible that such signals were of planetary nature \citep{bry13}, the expected flux dilution makes classification difficult. Thus, candidates with centroid offsets were removed from the DR25 candidate catalog. Employing a similar test for \emph{K2} would be difficult given the spacecraft's roll motion. The {\tt DAVE} algorithm \cite{kos19} considered each individual cadence (instead of the entire transit) to carry out a similar procedure for \emph{K2} and was able to identify 96 centroid offsets from the list of known candidates. While this method is helpful, it lacks the statistical strength of the original centroid test. Instead, we choose to leverage the \emph{Gaia} DR2 catalog to identify these sub-pixel background sources. Doing so, we are able to identify potential contaminates to within $1\arcsec$ of the target star \citep{zie18}, the equivalent limit of the \emph{Kepler} DR25 centroid offset test.\footnote{\emph{Gaia's} spatial resolution reduces to $2\arcsec$ for magnitude difference greater than 5.}  

\begin{figure}
\centering \includegraphics[width=\columnwidth{}]{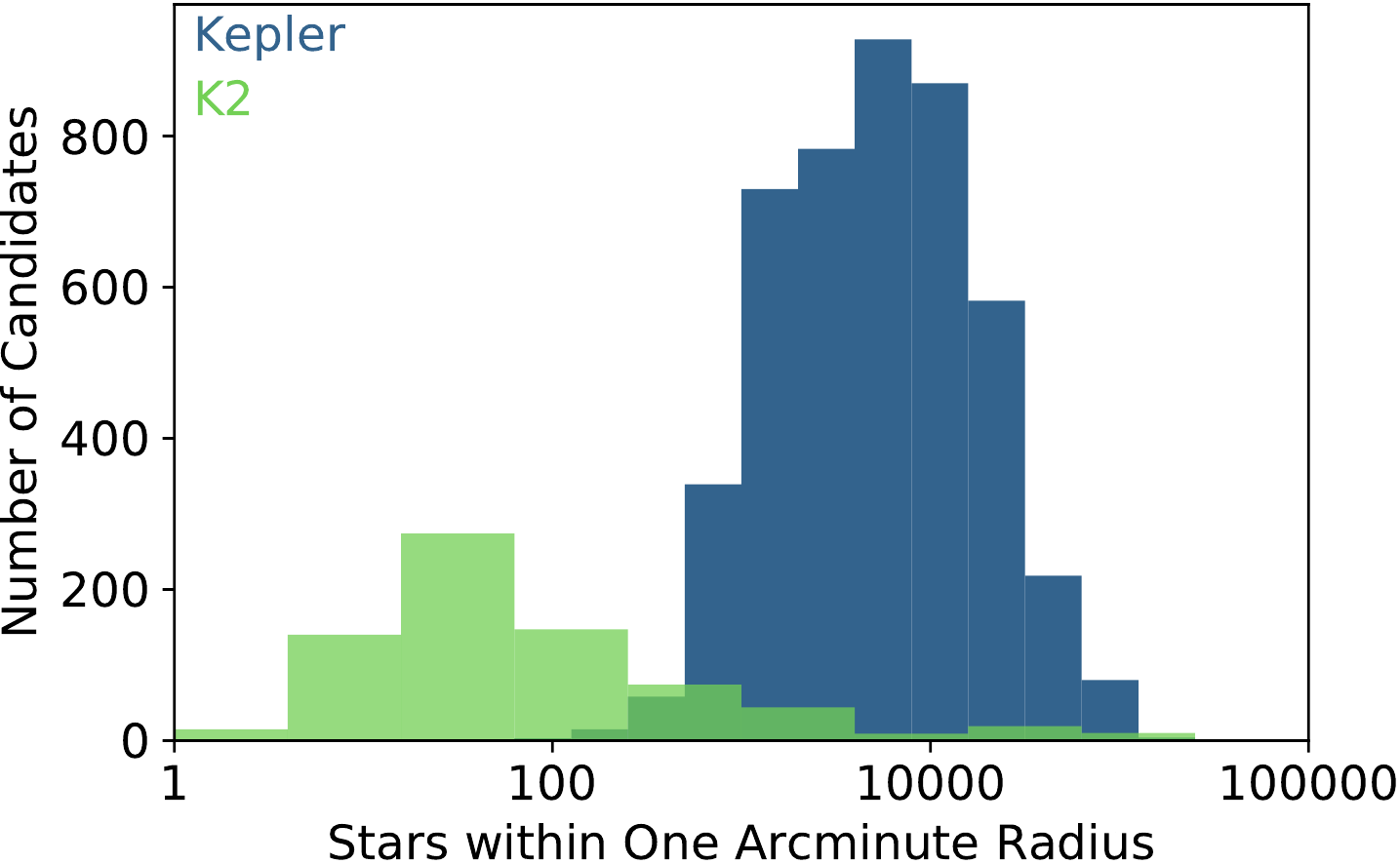}
\caption{The number of \emph{Gaia} DR2 sources within a one arcminute radius of each planet candidate. This plot shows that \emph{K2} candidate hosts (green) generally occupy more isolated fields than \emph{Kepler} candidate hosts (blue). A characteristic aligned with the fact that $92\%$ of our \emph{K2} candidates are observed at an absolute galactic latitudes greater than the edge of the \emph{Kepler} field ($\mid b \mid>22\degr$), where stellar density subsides.
\label{fig:seperation}}
\end{figure}

To estimate the rate of contamination in our sample we consider the \emph{Kepler} certified false positive (CFP) table \citep{bry17} alongside the \emph{Kepler} DR25 candidate catalog. The \emph{Kepler} CFPs represents a sample of 3,590 \emph{Kepler} signals not granted candidacy and thoroughly investigated to ensure a non-planetary origin. Carrying out all of our vetting procedures (including the contaminant identification from \emph{Gaia} DR2) to both the CFP table and the DR25 candidate list, we would expect to find 169 CFPs and 3,470 DR25 candidates (a $\sim4.6\%$ contamination rate). However, this rate is likely an upper limit. The astrophysical false positive rate for background EBs should recede with increased distance from the galactic plane, given the expected change in aperture crowding as a function of galactic latitude. Overall, the \emph{Kepler} target stars are closer to the galactic plane ($b \lesssim 20\degr$) compared to the \emph{K2} targets, which largely occupy a higher absolute galactic latitude ($\mid b \mid\gtrsim 20\degr$). This apparent density distinction is shown in Figure \ref{fig:seperation}, where we present the number of \emph{Gaia} sources within a one arcminute radius of each candidate host. The \emph{Kepler} candidates clearly occupy a more crowded field than a majority of our \emph{K2} candidates, subjecting \emph{Kepler} targets to a heightened occurrence of background contaminating sources. Overall, the CFP table is unique to the \emph{Kepler} prime mission and differences in instrument performance, catalog construction, and stellar fields limit our ability to make precise contamination rate estimates for our \emph{K2} catalog.

\begin{figure}
\centering \includegraphics[width=\columnwidth{}]{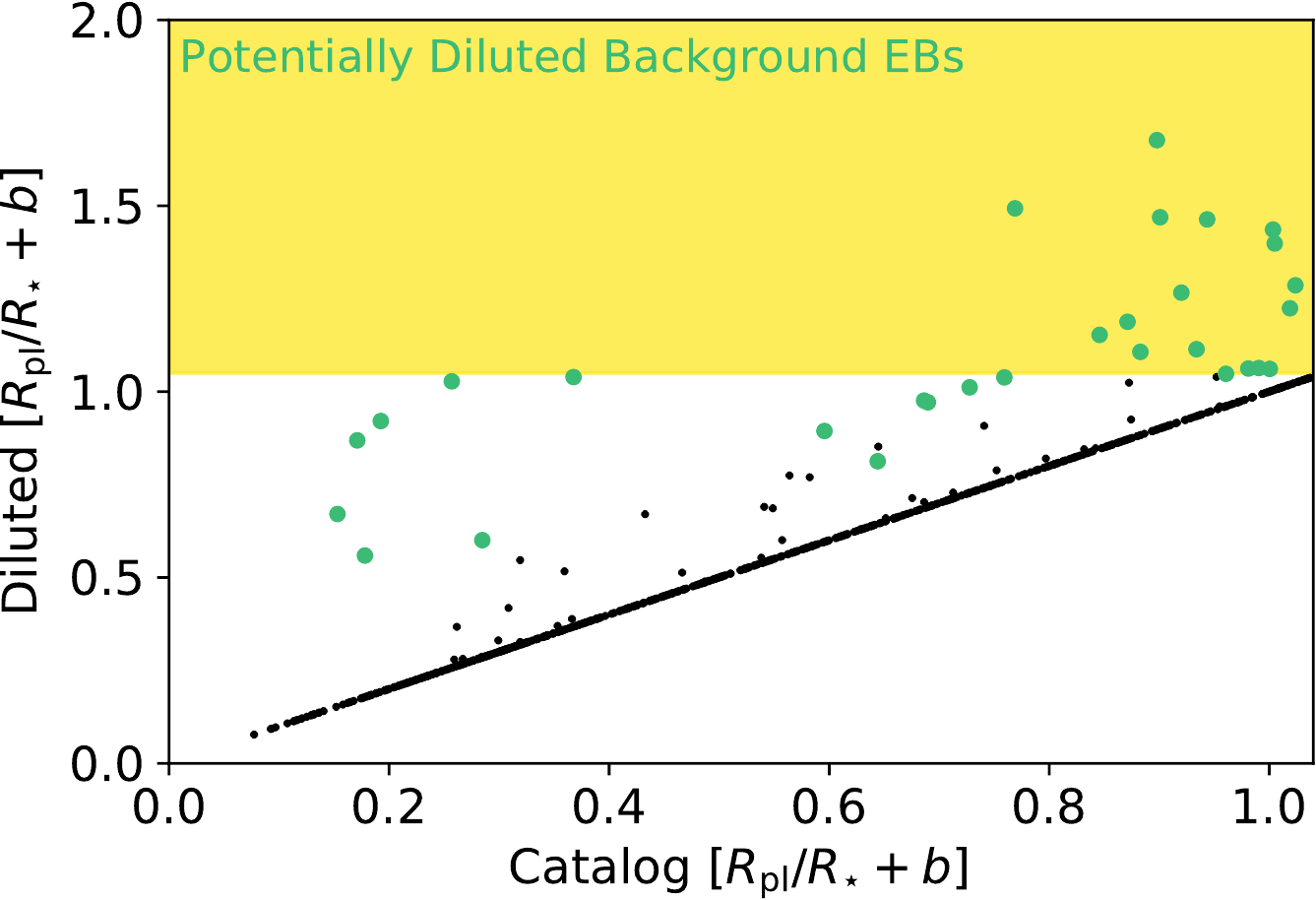}
\caption{The expected Equation \ref{eq:EB} values for the candidates within our catalog, conservatively assuming that the signals originated from the dimmest aperture-encased sources (Diluted). These quantities are plotted against the catalog values, which assume the signal host was the brightest aperture-encased source (Catalog). The yellow regions, which includes 18 candidates, indicate signals that could be astrophysical false positives ($R_{\mathrm{pl}}/R_{\star}+b>1.04$) with their transit signal diluted. We also found an additional 11 candidates that would have produced a $R_{\mathrm{pl}}/R_{\star}$ greater than 0.3 under this worse-case scenario assumption, presenting EB like properties and warranting rejection from our catalog. Signals that exceed either of these thresholds have been colored in green to indicate a potential background EB signal.
\label{fig:fluxDilution}}
\end{figure}

To further assess our ability to remove background EBs, we can perform a worst-case scenario test. The \emph{Gaia} DR2 source catalog is essentially complete \footnote{Visual companions, who are not spatially resolved, are an exception to this completeness claim.} for photometric $G$-band (comparable to the $Kepler$-band) sources brighter than 17 magnitudes, reducing to $\sim80\%$ completeness for $G=20$ targets \citep{bou20}. Correspondingly, in Figure \ref{fig:cdpp} it is shown that our CDPP values increase as a function of magnitude, reducing our pipeline's ability to detect signals around these dim targets. For example, a Kep.$=17$ star requires an occultation that exceeds $1\%$ to qualify as a TCE ($8.68\sigma$), increasing to $40\%$ for targets with Kep.$=20$. When these dim stars are within the aperture of a brighter source, the photometric noise floor is further elevated by the noise contribution from both stars. By combining the low probability of greater than $40\%$ eclipses and the $80\%$ stellar completeness at $G=20$, \emph{Gaia} provides sufficient coverage of the parameter space where background EBs could be hidden, yet still detectable by our pipeline. Thus, we can bound our contamination rate by making a conservative assumption that all of our candidate signals originate from the dimmest aperture-encased \emph{Gaia} DR2 identified source. Currently, our candidate catalog assumes each transit signal corresponds to the brightest star within the aperture, however, such an assumption may misidentify diluted background EBs. Our worse-case scenario test helps quantify the magnitude of this contamination and the corresponding results are displayed in Figure \ref{fig:fluxDilution}. We found 74 of our candidate targets contain additional aperture-encased \emph{Gaia} source. Nine of these targets exhibit transits that can only be physically explained by a signal from the brightest star. In other words, the transit brightness reduction exceeds that of the entire flux contribution from the background stars. We found 17 candidates that would exhibit $R_{\mathrm{pl}}/R_{\star}+b$ values greater than our catalog threshold (1.04) and an additional 13 candidates that would not meet our $R_{\mathrm{pl}}/R_{\star}/le0.3$ requirement. If we assume all 30 of these signals originated from a background source, we establish a background EB contamination rate of $4.0\%$. However, the true parent source of these candidates remains unclear, thus this estimation is again an upper limit. In addition, we acknowledge that our detection metrics could remain averted by small radius-ratio non-grazing EBs, but we expect such cases to be rare. Overall, we expect contamination from background EBs to make up less than 5\% of our candidate list.

\section{Occurrence Rate Recommendations}
\label{sec:suggest}
All exoplanet demographic analyses require defining an underlying stellar sample of interest. Our catalog used nearly the entire \emph{K2} target sample, which may be too broad for future analysis. The completeness measurements given here may not accurately reflect that of a reduced target list. Therefore, we recommend users select their own stellar sample and consult the  injection/recovery summary table to assess the completeness of their selected targets for the highest degree of accuracy.\footnote{\label{noteWeb}\url{www.jonzink.com/scalingk2.html}} However, the completeness parameters provided in Table \ref{tab:complete} are robust to minor sample selection modifications. Evidence of this claim is provided by the fact that the M dwarf and AFGK dwarf samples provide consistent results, see Figure \ref{fig:TCEcomplete}, despite known differences. 

In addition, 41,061 targets in our sample lacked stellar parameters, requiring the assumption of solar values. This may modify the results once these parameters become available. The remaining stellar parameters were provided by \cite{hub16} for 94,769 targets and \cite{har20} for 222,088 targets. While \cite{zin20b} showed the offset between these two methods is minimal, making this mixture of catalogs reasonable, a uniform stellar parameterization would provide more homogeneous results. Fortunately, the pipeline's stellar agnosticism makes parameter updating uncomplicated. As additional data from the forthcoming \emph{Gaia} DR3 becomes available, it will likely modify many of these stellar values and alter the underlying sample of dwarf stars. We suggest users implement the most up-to-date stellar parameters along with the injection/recovery summary table to evaluate sample completeness.

Upon close inspection, the \emph{K2} population appears more stochastic along the main sequence (see Figure \ref{fig:HR}). This is largely attributed to the guest observing selection process for the \emph{K2} fields, where individual proposals each applied their own target selection criteria to construct target lists that addressed their specific science goals. This could lead, for instance, to situations where the G dwarfs in a given campaign represent a distinct population from the G dwarfs in another campaign (e.g. probing different ranges of stellar metallicity, which is known to impact planet occurrence rates), or from a given population of field G dwarfs. \citet{zin20b} looked at the stellar population around Campaign 5 and found this latter selection effect did not provide a biased sample of FGK dwarfs for C5. However, for occurrence rate calculations, similar inspection of other campaigns should be carried out to ensure each campaign provides a uniform representation of their respective region of the sky. Where that does not appear to be true, users of this catalog are encouraged to independently select a set of targets from the full set of available targets (using, for instance, \emph{Gaia} properties) that more clearly represents an unbiased sample of the desired population.

Incorporating \emph{Gaia} DR2 into our pipeline enabled us to provide more accurate planet radii measurements. \citet{cia17} showed that non-transiting stellar multiplicity can artificially reduce transiting planet depths, leading to an overestimation in the occurrence of Earth-sized planets by 15-20\%. Our pipeline used the \emph{Gaia} DR2 to account for neighboring flux contamination, improving the precision of our radii measurements and the accuracy of future occurrence estimates. However, planet radius is markedly dependent on the underlying stellar radius measurements and our catalog is derived independent of such parameterization. Therefore, we did not impose any strict upper limits on planet radius and found 28 of our candidates have planet radii exceeding $30R_\Earth$. These candidates are likely astrophysical false positives. We suggest users consider an upper radius bound when carrying out occurrence analyses. Users may also consider using the \emph{Gaia} renormalized unit weight error (RUWE) values to further purify their sample of interest, as suggested by \citep{bel20}.  

Ideally, the reliability would also be updated as additional information on stellar and planetary parameters are made available. This is possible using the reliability summary table, but given the small number FAs found we expect very minor changes to occur.

It is important to note that our completeness measurements do not address the window function, which requires three transits occur within the available photometry. All injections were required to have at least three transits occurring within this window, removing this detection probability from the calculated completeness. In testing, we found most light curves follow the expected analytic probability formula ($prob$):

\begin{equation}
\label{eq:win}
\begin{aligned}[t]
prob & =1; & P<t_{\text{span}}/3\\
prob & =\frac{t_{\text{span}}}{P}-2;  & \frac{t_{\text{span}}}{3}\le P \le \frac{t_{\text{span}}}{2}\\
prob & =0;  & P> t_{\text{span}}/2,
\end{aligned}
\end{equation}
where $P$ is the signal period and $t_{\text{span}}$ is the total span of the data (see Figure 11 of \citealt{zin20a}). However, intra-campaign data gaps exist and should be carefully considered in any occurrence rate calculations.

\section{Summary}
\label{sec:summary}
We provide a catalog of transiting exoplanet candidates using \emph{K2} photometry from Campaign 1-8 and 10-18, derived using a fully automated detection pipeline. This demographic sample includes 747 unique planets, 366 of which were previously unidentified. Additionally, we found 57 multi-planet candidate systems, of which 18 are newly identified. These discovered systems include a K dwarf (EPIC 249559552) hosting two sub-Neptune candidates in a 5:2 mean-motion resonance, and an early-type F dwarf (EPIC 249731291) with two short period gas giant candidates, providing an interesting constraint on formation and migration mechanisms. Follow-up observations and validation of these and a number of the other new candidates presented in this catalog is currently underway (Christiansen et al. in prep).

We employed an automated detection routine to achieve this catalog, enabling measurements of sample completeness and reliability. By injecting artificial transit signals before {\tt EVEREST} pre-processing, we provide the most accurate measurements of \emph{K2} sample completeness. Additionally, we used the inverted light curves to measure our vetting software's ability to remove systematic false alarms from our catalog of planets, providing a quantitative assessment of sample reliability. Using this planet sample, and the corresponding completeness and reliability measurements, exoplanet occurrence rate calculations can now be performed using \emph{K2} planet candidates, which will be the subject of the next papers in the Scaling K2 series. With careful consideration of each data set's unique window functions, the \emph{Kepler} and \emph{K2} planet samples can now be combined, maximizing our ability to measure transiting planet occurrence rates throughout the local galaxy.

\section{Acknowledgements}

We thank the anonymous referee for their thoughtful feedback. This work made use of the gaia-kepler.fun crossmatch database created by Megan Bedell. The simulations described here were performed on the UCLA Hoffman2 shared computing cluster and using the resources provided by the Bhaumik Institute. This research has made use of the NASA Exoplanet Archive and the Exoplanet Follow-up Observation Program website, which are operated by the California Institute of Technology, under contract with the National Aeronautics and Space Administration under the Exoplanet Exploration Program. This paper includes data collected by the \emph{Kepler} mission and obtained from the MAST data archive at the Space Telescope Science Institute (STScI). Funding for the \emph{Kepler} mission is provided by the NASA Science Mission Directorate. STScI is operated by the Association of Universities for Research in Astronomy, Inc., under NASA contract NAS 5–26555. J. Z. acknowledges funding from NASA ADAP grant 443820HN21811. K. H-U and J. C. acknowledge funding from NASA ADAP grant 80NSSC18K0431.

\software{{\tt EVEREST} \citep{lug16,lug18}, {\tt TERRA} \citep{pet13b}, {\tt EDI-Vetter} \citep{zin20a}, {\tt PyMC3} \citep{sal15}, {\tt Exoplanet} \citep{for19}, {\tt RoboVetter} \citep{tho18}, {\tt batman} \cite{kre15}, {\tt emcee} \citep{for13}}


\bibliography{paper}{}
\bibliographystyle{aasjournal}



\end{document}